\documentclass[times,review,preprint,authoryear]{elsarticle}
\usepackage{medima}
\usepackage{framed,multirow}
\usepackage{amsmath}

\usepackage{amssymb}
\usepackage{latexsym}

\usepackage{url}
\usepackage{multirow}
\usepackage[table,xcdraw]{xcolor}
\usepackage{amssymb}
\usepackage{pifont}
\newcommand{\cmark}{\ding{51}}%
\newcommand{\xmark}{\ding{55}}%
\usepackage{graphicx}

\usepackage{hyperref}

\definecolor{newcolor}{rgb}{.8,.349,.1}

\journal{Preprint submitted to Medical Image Analysis}

\begin{document}

\verso{Jin Zhu \textit{et~al.}}

\begin{frontmatter}

\title{A residual dense vision transformer for medical image super-resolution with segmentation-based perceptual loss fine-tuning}%

\author[1]{Jin \snm{Zhu}\corref{cor1}}
\cortext[cor1]{Corresponding author: zhujin1121@gmail.com; g.yang@imperial.ac.uk}
\author[2,3]{Yang \snm{Guang}\corref{cor1}\fnref{fn1}}

\author[1]{Pietro \snm{Li\'{o}}\fnref{fn1}}
\fntext[fn1]{Co-last senior authors contributed equally.}

\address[1]{Department of Computer Science and Technology, University of Cambridge, Cambridge CB3 0FD, UK}
\address[2]{National Heart and Lung Institute, Imperial College London, London SW3 6LY, UK}
\address[3]{Cardiovascular Research Centre, Royal Brompton Hospital, London SW3 6LR, UK}

\received{06 March 2023}
\finalform{None}
\accepted{None}
\availableonline{None}
\communicated{None}

\begin{abstract}
Super-resolution plays an essential role in medical imaging because it provides an alternative way to achieve high spatial resolutions and image quality with no extra acquisition costs. In the past few decades, the rapid development of deep neural networks has promoted super-resolution performance with novel network architectures, loss functions and evaluation metrics. Specifically, vision transformers dominate a broad range of computer vision tasks, but challenges still exist when applying them to low-level medical image processing tasks. This paper proposes an efficient vision transformer with residual dense connections and local feature fusion to achieve efficient single-image super-resolution (SISR) of medical modalities. Moreover, we implement a general-purpose perceptual loss with manual control for image quality improvements of desired aspects by incorporating prior knowledge of medical image segmentation. Compared with state-of-the-art methods on four public medical image datasets, the proposed method achieves the best PSNR scores of 6 modalities among seven modalities. It leads to an average improvement of $+0.09$ dB PSNR with only 38\% parameters of SwinIR. On the other hand, the segmentation-based perceptual loss increases $+0.14$ dB PSNR on average for SOTA methods, including CNNs and vision transformers. Additionally, we conduct comprehensive ablation studies to discuss potential factors for the superior performance of vision transformers over CNNs and the impacts of network and loss function components. The code will be released on GitHub with the paper published.
\end{abstract}

\end{frontmatter}


\section{Introduction}\label{sec:intro}
\noindent Medical images are crucial in the current clinical process, including early detection, staging, guiding intervention procedures and surgeries, radiation therapy, and monitoring disease recurrence [\cite{MedicalImaging2013Nature}]. For example, computed tomography (CT) and magnetic resonance (MR) scans are widely used in the diagnosis and study of Alzheimer's disease [\cite{ADDiagnosis}], stroke [\cite{Stroke}], autism [\cite{Autism}], Parkinson's disease [\cite{Parkinson}] and COVID-19 [\cite{ReviewCOVID}]. Although people have witnessed the importance of in-vivo radiology images, their spatial resolution is subject to the scan time, body motion, dose limit and hardware configurations. Thus, super-resolution methods are introduced as alternative post-processing to achieve higher resolution and better image quality without extra acquisition costs [\cite{MedSR2020Review}]. 
\\
\\
Technically, image super-resolution is a process to recover an image of high-resolution (HR) from low-resolution (LR) versions. Depending on the number of input and output images, we briefly divide the methods into two groups: single-image super-resolution (SISR) and multi-image super-resolution. With the rapid development of deep learning algorithms, SISR methods with neural networks achieve superior performance on natural images than the previous interpolation-based, reconstruction-based, and learning-based methods [\cite{SRReview18, SRReview, SRReview2022Lepcha}]. However, introducing deep neural networks to medical image super-resolution tasks is still an open problem. In the clinic, super-resolved images always proceed to medical image analysis tasks, and the datasets are relatively small [\cite{Shen2017Review, MIA2017Review, deepLearning2020chan}]. Thus, super-resolution methods for medical images require novel mechanisms and modifications on training datasets, loss functions, evaluation metrics and network architecture design to preserve sensitive information and to enhance the structures of interest for radiologists and physicians [\cite{MedSR2020Review}].
\\
\\
Recently, convolutional neural networks (CNNs) and generative adversarial networks (GANs) have successfully applied to medical image super-resolution tasks [\cite{MedSR2020Review}]. However, it is still an open problem to involve vision transformers (ViTs), which achieve state-of-the-art performance on a wide range of natural image restoration tasks [\cite{ViT2022ReviewHan}] and medical image analysis tasks [\cite{ViTReviewOnMI2022Henry, ViTReviewonMIA2022Shen}]. Exploring and discussing the robustness, capacity, efficiency and limitation of vision transformers on medical image SR tasks is necessary. On the other hand, acknowledged tricks of CNN architecture design, such as localisation operation, residual connections and feature fusion, are worth introducing to CNN-ViT hybrid models for potential performance improvement. For example, inspired by the shared weights and localisation operations of CNNs, a shifted window vision transformer (Swin Transformer [\cite{swinTransformer}]) is proposed for high-level image processing tasks. The novel Swin layers are then applied in natural image restoration [\cite{swinir, fan2022sunetDenoising}] and segmentation [\cite{he2022swinSegmentation, cao2021swin_seg_MI}]. 
\\
\\
Additionally, prior knowledge of related medical image tasks, such as segmentation, can benefit the upstream super-resolution task. On the one hand, the challenges of performing image quality assessment (IQA) on enhanced images [\cite{chandler2013sevenIQA}] and on medical images [\cite{chow2016MedicalIQA}] still exist. Generally, IQA of generated and super-resolved natural images mainly includes reconstruction accuracy and human perception. Since current SISR methods are getting close to the limitation of signal fidelity metrics [\cite{wang2009MSE}], perceptual quality assessment methods [\cite{zhai2020perceptualIQAReview}] have become more and more critical. In contrast, various artefacts in medical images, mainly caused by the hardware of imaging systems and the body motion of individuals, are never seen in natural images. Peak signal-to-noise ratio (PSNR) and structure similarity (SSIM [\cite{SSIM}]) are prevalent in almost every medical image SR work. However, directly and only using the IQA methods designed for natural images may not be reliable in medical image SR tasks. In a supplemental manner, researchers evaluate the quality of SR images with the performance of downstream medical image analysis tasks such as segmentation [\cite{xia2021CMRSR}]. Although the quality measurement of medical images does not equal diagnostic accuracy [\cite{DiagnosticQA2010IEEE}], radiologists and medical consultants always prefer high-quality images for accurate diagnosis. In addition to the measurement of machine perception, the prior knowledge of pre-trained segmentation models can also benefit the training of medical image super-resolution models, similar to the existing perceptual losses [\cite{PerceptualLoss, ESRGANPercptualLoss}].
\\
\\
Thus, we explore the possibility of extending successful architectures in CNNs to vision transformers to improve the single-image super-resolution performance of medical images efficiently and robustly. We propose a Residual Dense Swin Transformer (RDST) as a novel backbone for SR tasks by introducing residual dense connections [\cite{DenseNet, SRDenseNet}] and local feature fusion [\cite{RDN, ESRGAN}] to SOTA vision transformers. Meanwhile, we take segmentation as a typical medical image analysis task in the clinic and connect it with the upstream super-resolution task for model training and result evaluation. We present a perceptual loss based on the prior knowledge of the pre-trained segmentation U-Net [\cite{UNet}] and extend its variants to a wide range of SOTA SISR models, including CNNs and ViTs. In this work, we focus on supervised super-resolution with a single magnification scale (i.e. $\times 4$). At the same time, the proposed method has potential applicability to semi-/un-supervised SR tasks with multi or arbitrary magnification scales. For a comprehensive comparison with SOTA SISR methods, we run experiments on four big and small public medical image datasets, including brain MR images, cardiac MR images and CT scans of COVID patients. Ablation studies are also designed to discuss the impacts of critical characteristics of the proposed model architecture, perceptual loss and training tricks. 
\\
\\
The paper is organised as follows: in Section \ref{sec:background}, we briefly summarise related works of single image super-resolution; in Section \ref{sec:method}, we introduce the proposed residual dense vision transformer and the segmentation task based perceptual loss; in Section \ref{sec:experiments}, we describe the experiment settings; in Section \ref{sec:result}, we illustrate the qualitative and quantitative results and discuss the essential characteristics of the proposed method in contrast with SOTA SISR methods; and in Section \ref{sec:conclusion}, we provide concluding remarks of this work.
\\
\section{Related work} \label{sec:background}
\noindent This section provides a comprehensive review of advanced super-resolution networks with medical image applications.  

\subsection{Super-resolution networks}
\noindent Implementation of super-resolution networks includes CNNs [\cite{SRCNN, EDSR, RCAN}], GANs [\cite{SRGAN, ESRGAN}],  vision transformers [\cite{zamir2022restormer, swinir, lu2022ESRT}], diffusion models [\cite{IPT, li2022srdiff, ho2022cascaded_diffusion_model_1}] and hybrid methods [\cite{gao2022LBNet}]. These frameworks involve a wide range of deep learning techniques, such as recursive learning [\cite{tai2017memnet, DRCN, tai2017DRRN, gao2022LBNet}], local and global residual learning [\cite{RDN, li2018multi_scale_residual_sr, hui2018IDN, VDSR}], multi-path learning [\cite{EDSR, han2018DSRN, ren2017CNF, mehri2021mprnet}], attention mechanisms [\cite{RCAN, dai2019SAN, HAN}] and U-Net architectures [\cite{UNetSR, liu2018unet_image_restoration, qiu2022UNetSR_medical_CardiacMR}].  For a comprehensive review of SISR networks, we refer to the three citations [\cite{SRReview, SRReview18, SRReview2022Lepcha}].
\\
\\
Early SISR networks such as SRCNN [\cite{SRCNN}], VDSR [\cite{VDSR}], MemNet\cite{tai2017memnet}] and DRCN [\cite{DRCN}] follow the pre-upsampling architecture design, which first upsamples the LR image to the desired size and refines the magnified results with limited convolution layers. Then, post-upsampling frameworks are proposed in ESPCN [\cite{SubPixel}] and FSRCNN [\cite{dong2016FSRCNN}], which apply feature learning in low-resolution space and achieve feature maps magnification subsequently. Moreover, residual connections are widely used for advanced representation capability in deep SR networks without gradient vanishing. SRResNet [\cite{SRGAN}] first introduces the residual block of ResNet [\cite{ResNet}] with GANs for photo-realistic image generation. Then, EDSR [\cite{EDSR}] and SRDenseNet [\cite{SRDenseNet}] further improve the block architectures by removing the unnecessary batch-normalisation layer [\cite{batch_normal}] and introducing dense connections. Moreover, RDN [\cite{RDN}] and ESRGAN [\cite{ESRGAN}] implement local feature fusion to reduce the computation cost of dense blocks and encourage more stable information and gradient flows.
\\
\\
Attention mechanism adaptively realises deep neural networks to most informative regions of the input, leading to a more efficient and effective understanding of complex scenes in computer vision tasks [\cite{guo2022attention_review_cv}], such as super-resolution [\cite{zhu2021attention_sr_review}]. RCAN [\cite{RCAN}] first introduce the channel attention mechanism into residual SR networks, leading to flexible processing of low-/high-frequency information with channel-wise interdependence learning on features maps. Additionally, advanced CNN-based methods apply multi-scale Laplacian pyramid attention [\cite{anwar2020dense_resdiual_lapSR}] second-order channel attention [\cite{dai2019SAN}] and spatial attention [\cite{choi2017deep_selnet, liu2020residual_spatial_attention_sr}]. HAN [\cite{HAN}] further implements a holistic attention network with the hybrid channel-spatial attention and the layer attention modules. 
\\
\\
Transformers are first applied for natural language processing [\cite{vaswani2017attention_is_all_you_need}] and then introduced to vision tasks [\cite{ViT2022ReviewHan, ViTReviewOnMI2022Henry}] by embedding image and feature maps to tokens. For low-level image restoration tasks, IPT [\cite{IPT}] involves CNN-based shallow feature embedding before self-attention layers and transfer learning of pre-trained model on ImageNet [\cite{deng2009imagenet}]. SwinIR [\cite{swinir}] proposes an efficient transformer backbone based on the shifted-window attention [\cite{swinTransformer}] and achieves SOTA performance on a broad range of natural image restoration tasks. Similarly, Uformer [\cite{wang2022uformer}] implements a U-Net framework with locally-enhanced window transformer blocks and a novel multi-scale restoration modulator for multi-task learning on image restoration. Both SwinIR and Uformer transformers successfully avoid the expensive computational cost of global self-attention on high-resolution feature maps and the limitation of transformers in capturing local dependencies. Meanwhile, the multi-deconv transposed attention and gated-dconv feed-forward network are proposed in Restormer [\cite{zamir2022restormer}] to aggregate local and non-local pixel interactions and perform controlled feature transformation. Additionally, hybrid CNN and transformer methods are proposed for efficient and lightweight super-resolution [\cite{lu2022ESRT, zou2022self_lightweight_sr, fang2022hybrid_cnn_transformer_lightweight_image_sr}].

\subsection{SR on medical images}
\noindent Before the explosion of deep neural networks, early applications of medical image super-resolution mainly relied on interpolation, reconstruction and example-learning methods [\cite{isaac2015mi_sr_review_early}]. Although these methods involve multi frames [\cite{peled2001superresolution_mr, greenspan2002mri_sr, shilling2008super_sr}] and reference slices [\cite{rousseau2008brain_sr, manjon2010mri_sr}] for HR image reconstruction, they achieve poor performance due to the limited representational capacity and lack of additional information of the training data. 
\\
\\
Researchers extend SRCNN [\cite{SRCNN}] with proper modifications to brain MR super-resolution [\cite{pham2017brain_mr_3d, mcdonagh2017context, du2020sr_3d_mr}] and cardiac MR reconstruction with multi-input [\cite{oktay2016multi_input_cardiac_image_sr}]. [\cite{shi2018mr_sr_wide_resnet} presents a pre-upsampling SR framework modified residual blocks of EDSR [\cite{EDSR}] for 2D brain MR slices. [\cite{chen2018brain_sr_3d_densenet} and [\cite{li2021volumenet_mr_ct_3d_sr}] applies 3D convolution in one dense connected block [\cite{SRDenseNet}] for 3D brain MR super-resolution and liver tumour CT images. In [\cite{zhao2019channel_CSN_mr_super_resolution}], researchers propose a channel splitting network with global feature fusion [\cite{RDN} and merge-and-run mapping [\cite{zhao2016deep_merge_and_run_mappings}], leading to a representational redundancy decline and hierarchical features integration. U-Net architectures are also widely used in medical image super-resolution tasks [\cite{UNetSR, qiu2022UNetSR_medical_CardiacMR, qiu2021progressive_unet_Covid_ct_sr}], especially for multi-task learning with segmentation [\cite{SegSR, bhandary2022double_unet_sr_seg_cell}]. 
\\
\\
Adversarial learning and the perceptual loss [\cite{PerceptualLoss}] are also widespread in single slice super-resolution due to the expensive computation cost of 3D operations [\cite{3DSR, sanchez2018brain_3D_srgan}]. [\cite{gu2020medsrgan}] implements a conditional GAN framework with a modified RCAN [\cite{RCAN}] for 2D super-resolution on CT and MR images. FA-GAN [\cite{jiang2021fagan}] proposes a fused attentive GAN with CNN-based channel and non-local attentions for 2D MR super-resolution. FP-GANs [\cite{you2022fine_gan_brain_mr_sr_wavelet_domain}] conducts SR in a divide-and-conquer manner with multiple ESRGAN [\cite{ESRGAN}] in the wavelet domain. CycleGAN [\cite{CycleGAN}] also benefits medical image super-resolution [\cite{CycleGANSR}], such as with lesion-focused [\cite{de2021impact_cycle_gan_with_lesion_sr_}] and semi-supervised training [\cite{jiang2020novel_semi_sr_CT_cycle_gan}]. Meanwhile, Wasserstein distances [\cite{WGAN, WGANGP}] are also widely used [\cite{lyu2018mr_sr_wgan, shahidi2021breast, yang2018low_dose_ct_denoising_wgan, lyu2020multi, LFSR, MSGAN, MIASSR}]. 
\\
\\
Vision transformers boost the performance of medical image processing tasks such as reconstruction [\cite{guo2022reconformer, huang2022swin}, denoising [\cite{jang2022pet_denoising_restormer}] and segmentation [\cite{tang2022self}]. [\cite{puttaguntaa2022swinir_medical_sr}] present a SwinIR application on medical images, including chest x-ray, skin lesions and funds images. Meanwhile, [\cite{li2022swinir_multi_contrast_MRI_SR}] apply SwinIR for multi-contrast MR super-resolution with multi-scale contextual matching and aggregation schemes. 
\\
\\
Based on a comparison study of state-of-the-art single-image super-resolution methods on four public medical image datasets, weclaim that the main contributions of this work include:
\begin{itemize}
    \item A Residual Dense Swin Transformer (RDST) is proposed by introducing the residual dense connections to vision transformers. In the $\times 4$ SR experiments on four medical image datasets, it achieves the best PSNR scores of 6 modalities among 7 modalities in total. It leads to +0.09 dB PSNR improvement on average than the SOTA SISR method SwinIR with only 38\% parameters. Additionally, the SR results of RDST achieve the best segmentation accuracy of 8 sub-regions among all 15 target regions in the downstream segmentation tasks and increase the dice coefficient by 0.0029 on average than the SR results of SwinIR. 
    
    \item The lite version RDST-E further improves the model efficiency with hyper-parameter modification. It achieves comparable performance with the SOTA method SwinIR on both SR image quality (+0.06 dB PSNR on average of 7 medical image modalities) and downstream segmentation accuracy (-0.0026 dice coefficient on average of 15 target regions) but has only 20\% parameters of SwinIR and is 46\% faster than SwinIR on inference. 
    
    \item Two variants of SR perceptual loss are proposed with pre-trained segmentation U-Nets, dramatically improving the SR image quality by transferring prior knowledge of medical images in segmentation tasks to super-resolution tasks. The proposed losses successfully extend to various SOTA SISR methods, including CNNs and ViTs. Compared with the native L1 loss, the novel loss variant for SR fidelity (i.e. $\mathcal{L}_{E(1)}$) results in a noticeable improvement of +0.14 dB PSNR on average, while the proposed loss variant for machine perception (i.e. $\mathcal{L}_{HRL}$) leads to an improvement of +0.0023 dice coefficient on average in the downstream segmentation task.
\end{itemize}

\section{Methods}\label{sec:method}
\begin{figure*}[tp]
\centering
\includegraphics[width=0.9\textwidth]{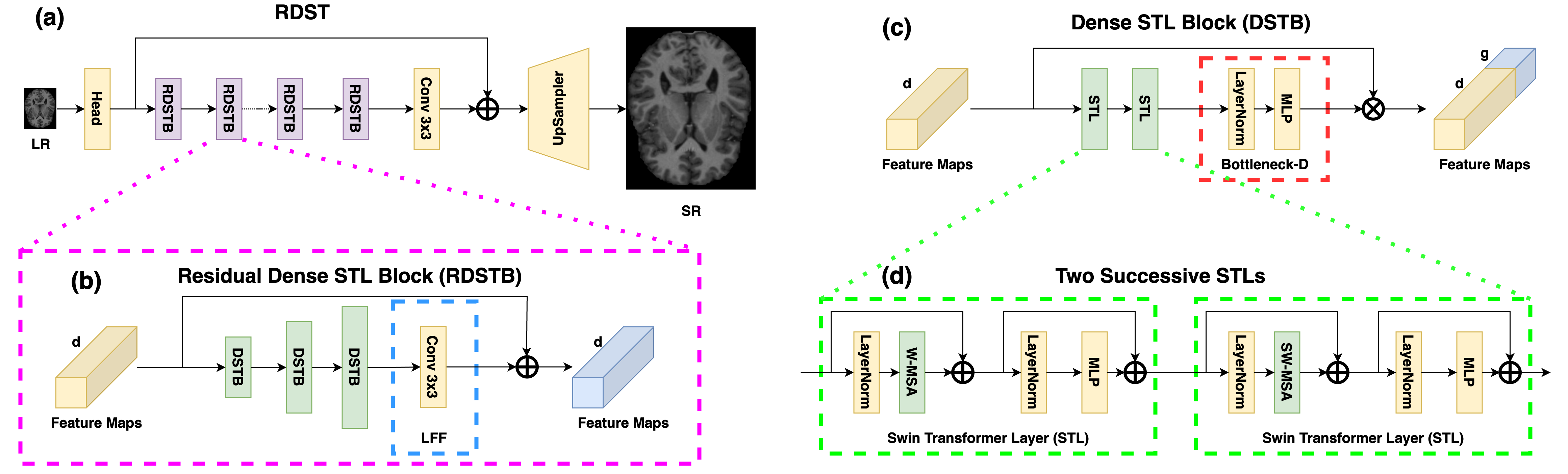}
\caption{Framework of the proposed RDST network. (a): the proposed RDST (notation represented with Eq. \ref{eqt:rdst_rdst}) consists of a convolution layer head for shallow feature extraction, a SubPixel-based UpSampler, $N$ RDSTB modules and a global residual connection. (b): A residual dense Swin transformer block (notation presented with Eq. \ref{eqt:rdst_rdstb}) is composed of three DSTB modules and a local feature fusion module (LFF), which compress the feature maps from $(3 \times g + d)$ to $d$. (c): A Dense STL Block (notation presented with Eq. \ref{eqt:rdst_dstb}) is composed of two successive swin transformer layers (STLs), a bottleneck layer and a concatenation operator. (d): The two successive shifted-window transformer layers (notation presented in Eq. \ref{eqt:rdst_stl}). Each STL consists of two-layer normalisation layers, one multi-head self-attention layer with regular or shifted windowing configurations (W-MSA and SW-MSA), one MLP layer and skip connections.
The patch embedding and un-embedding operations are ignored for a brief illustration. They convert feature maps from $[N \times d \times H \times W]$ to $[N_{w} \times P \times d]$ and vice visa to adjust linear layers and the convolution layers.}
\label{fig:rdst_net}
\end{figure*}

\noindent In this work, we mainly focus on single image super-resolution tasks with certain magnification scales (e.g. $\times 4$), which can be represented as:

\begin{equation} \label{eqt:rdst_lr_2_sr}
    I_{sr} = G_{s}(I_{lr}; \theta_{G}),
\end{equation}

\noindent where $s$ is the magnification scale, $I_{lr}$ and $I_{sr}$ are a pair of one input image with a low resolution of $[H \times W \times C]$ and its super-resolved output with a high resolution of $[sH \times sW \times C]$. Following the most popular and successful architecture of SISR networks, the proposed residual dense transformer consists of three components: a convolutional layer consisted shallow feature extraction head $\mathcal{H}$, a feature map up-sampler $\mathcal{U}$ and a main body for deep feature extraction (Fig. \ref{fig:rdst_net}-a). Mathematically, it can be represented as:

\begin{align}
    &F_{lr} = \mathcal{H}(I_{lr}), \nonumber \\
&F_{d} = F_{lr} + C_{k\times k}(\mathcal{R}^{n}(F_{lr})), \nonumber \\
&I_{sr} = \mathcal{U}_{s}(F_{d}), \label{eqt:rdst_rdst}
\end{align}

\noindent where $s$ is the magnification scale, $d$ is the basic dimension of feature map embedding, $C_{k \times k}$ is a convolutional layer with kernel size $k$ and $\mathcal{R}^{n}$ indicates $n$ sequentially stacked blocks. Similar to SOTA SISR methods, we use one $3\times 3 \times d$ convolutional layer as the head to extract the low-resolution feature maps $F_{lr} \in \mathbb{R}^{H \times W \times d}$ and a Sub-Pixel [\cite{SubPixel}] module as the up-sampler for super-resolution image generation. Regarding the main body for deep feature extraction, we use a skip connection for global residual learning and propose a residual dense swin transformer block (RDSTB), which will be introduced in detail in the following part.

\subsection{Residual dense swin transformer block}\label{sec:method_rdstb}
\noindent Shifted-windows transformer layer (STL) is used as the most basic unit in the proposed residual dense swin transformer block. To reduce the computation cost in the vision transformer, it splits the input feature maps of size $[H \times W]$ to windows of size $[M \times M]$ first and then applies standard multi-head self-attention localised in each window. To connect these local windows, in two successive STLs (Fig. \ref{fig:rdst_net}-d), the first STL applies regular window partition from top-left, while the second STL shifts the feature maps by $(\frac{M}{2}, \frac{M}{2})$ pixels before partition:

\begin{align}
    &\hat{F}^{i} = \mathcal{A_{W}}(\mathcal{N}(F^{i-1}))+F^{i-1}, \nonumber \\
&F^{i} = \mathcal{M}(\mathcal{N}(\hat{F}^{i})) + \hat{F}^{i}, \nonumber \\
&\hat{F}^{i+1} = \mathcal{A_{S}}(\mathcal{N}(F^{i}))+F^{i}, \nonumber \\
&F^{i+1} = \mathcal{M}(\mathcal{N}(\hat{F}^{i+1})) + \hat{F}^{i+1}, \label{eqt:rdst_stl}
\end{align}

\noindent where $\mathcal{A_{W}}$ and $\mathcal{A_{S}}$ are multi-head self-attention layers with regular and shifted window configurations, respectively. $\mathcal{N}$ is layer normalisation, and $\mathcal{M}$ is a multi-layer perceptron (MLP) consisting of two fully-connected layers with GELU non-linearity [\cite{GELUs}] in between. Skip connections with pixel-wise addition are applied after each module. The key advantages of STL are the localisation operation and shared weights, just like convolutional layers. Additionally, with the reshaping operation, the size of its output feature maps remains the same as the input feature maps (i.e. $[N \times d \times H \times W]$). Thus, it behaves like a convolutional layer, and successful designs in CNN-based SISR models can be easily introduced in STL-based models. 
\\
\\
First, we introduce dense connection [\cite{DenseNet, SRDenseNet}] to STLs. As shown in Fig. \ref{fig:rdst_net}-c, a Dense STL Block (DSTB) consists of two successive STLs $\mathcal{S}^{2}$ and an MLP-based bottleneck module $\mathcal{N}_{d\rightarrow g}$. Before concatenating to the input feature maps $F^{i-1}_{d}$, the new feature maps are compressed from $[N \times H \times W \times d]$ to $[N \times H \times W \times g]$ to reduce the computation cost further. Thus, the output of the $i$-th DSTB is computed as:

\begin{equation}
F^{i}_{d+g} = cat[F^{i-1}_{d}, \mathcal{B}_{d \rightarrow g}(\mathcal{S}^{2}(F^{i-1_{d}}))]. \label{eqt:rdst_dstb}
\end{equation}

\noindent Then, the residual dense swin transformer block (RDSTB, Fig. \ref{fig:rdst_net}-b) is proposed by applying local feature fusion (LLF) [\cite{RDN}] after stacking several DSTBs. One RDSTB consists of three successive DSTBs and a $3 \times 3$ convolutional layer for local feature fusion. As reported in SRDenseNet [\cite{SRDenseNet}] and [RDN \cite{RDN}], combining dense connections and LLF can preserve the feed-forward nature and extract local features without high computational costs and training problems. The convolution-based LLF controls the output information by reducing the number of feature maps from $(3 \times g + d)$ to $d$. As a result, the output feature maps $F^{i}$ of the $i$-th RDSTB block remains the same shape $[N \times d \times H \times W]$ as its input feature maps $F^{i-1}$:
\begin{equation}\label{eqt:rdst_rdstb}
    F^{i}_{d} = F^{i-1}_{d}+\mathcal{B}_{3\times g+d\rightarrow d}(\mathcal{D}^{3}(F^{i-1}_{d})), 
\end{equation}
where $\mathcal{D}^{3}$ is a group of three successive DSTBs and $\mathcal{B}$ is the bottleneck module for local feature fusion.

\begin{figure*}[t]
    \centering
    \includegraphics[width=0.8\textwidth]{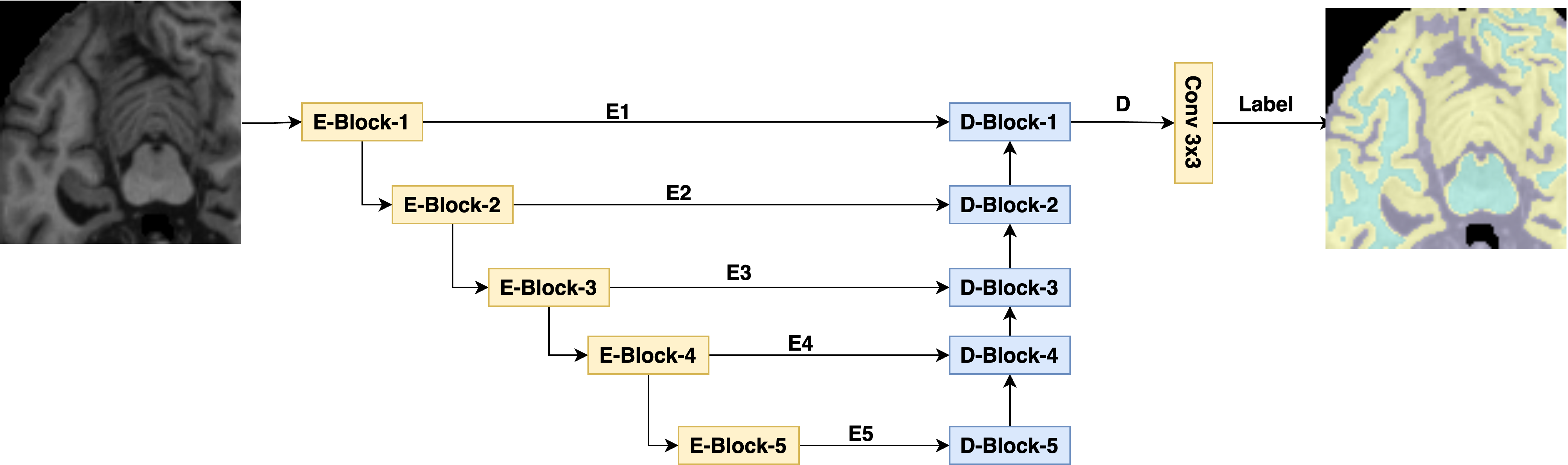}
    \caption{The segmentation U-Net [\cite{UNet}] is used in this work for two purposes: for perceptual losses and for segmentation-based SR evaluation. It consists of 5 levels of ResNet-based encoders and decoders, which are paired with skip connections. E1 to E5 indicate the output feature maps of each encoder block correspondingly, while D denotes the output of the last decoder.}
    \label{fig:rdst_unet_segloss}
\end{figure*}

\subsection{Segmentation U-Net based perceptual loss}\label{sec:mtd_unet}
\noindent The proposed method RDST is trained in two stages:  basic training with $\mathcal{L}_{1}$ loss and a fine-tuning stage with perceptual loss. In the first stage, the parameters are optimised by minimising the native pixel-wise L1 distance between the output SR images and HR ground truth images:
\begin{equation}
    \mathcal{L}_{1}(G(I_{lr}), I_{hr}) = \frac{1}{sH \times sW \times C}\left\| G(I_{lr}) - I_{hr} \right\|_{1},
\end{equation}
where $G$ is a RDST model, $s$ is the magnification scale, $I_{lr}$ is the input low resolution image with shape $[H \times W \times C]$ and $I_{hr}$ is the corresponding ground truth high resolution image. 
\\
\\
In the second stage, a U-Net [\cite{UNet}] based segmentation loss is proposed to fine-tune the parameters of the RDST after stage 1. Depending on the dataset, a U-Net model has been first trained for medical image segmentation with the same training data $I_{hr} \in \mathbb{R}^{sH \times sW \times C}$ and the corresponding segmentation labels $L_{hr}$:
\begin{equation}
\hat{\theta_{U}}=arg \; \underset{\theta_{U}}{\mathrm{min}} \:\: \mathcal{L}_{seg}(U(I_{hr}), L_{hr}), \label{eqt:rdst_unet_dice}
\end{equation}

\noindent where $U$ is the segmentation model, $\theta_{U}$ represents its trainable parameters and $\mathcal{L}_{seg}$ is a loss function for segmentation tasks.
\\
\\
Inspired by previous work on perceptual losses for SISR tasks [\cite{PerceptualLoss, SRGAN, ESRGAN}], we define the segmentation-based perceptual loss between two images $(I_{sr}, I_{hr})$ as the L1 distance between their feature maps of certain layers in the pre-trained U-Net. Depending on which layer to be used in the U-Net (Fig. \ref{fig:rdst_unet_segloss}), the segmentation-based perceptual loss of $(I_{sr}, I_{hr})$ can be represented as:

\begin{align}
    &\mathcal{L}_{E(i)} = \mathcal{L}_{1}(U[E_{i}](G(I_{lr}))-U[E_{i}](I_{hr})), \label{eqt:rdst_seg_loss_ei}\\
&\mathcal{L}_{D} = \mathcal{L}_{1}(U[D](G(I_{lr}))-U[D](I_{hr})), \label{eqt:rdst_seg_loss_d}
\end{align}
where $U[E_{i}]$ indicates the $i$-th block of the encoder and $U[D]$ is the decoder. Furthermore, the perceptual loss can also be defined with the similarity of predicted segmentation labels of $I_{sr}$ and $I_{hr}$:

\begin{equation}
    \mathcal{L}_{HRL} = 1 - \mathcal{L}_{seg}(U(I_{sr}), U(I_{hr})). \label{eqt:rdst_seg_loss_dice}
\end{equation}
In this work, we use dice coefficient [\cite{diceloss}] as $\mathcal{L}_{seg}$ to evaluate the distance between two binary segmentation labels $X$ and $Y$, because it is popularly used in medical image segmentation tasks [\cite{ma2021Seg_loss_review_MIA}]:
\begin{equation}
    Dice(X, Y) = \frac{2\left| X\bigcap Y \right|}{\left| X \right|+\left| Y\right|}. \label{eqt:rdst_dice_coefficient}
\end{equation}

\noindent During model training and fine-tuning, these segmentation-based perceptual loss variants can be used with the native $\mathcal{L}_1$ loss:
\begin{equation}\label{eqt:rdst_loss_finetune}
    \mathcal{L}_{SR} = \alpha \mathcal{L}_{1} + \lambda \mathcal{L}_{U},
\end{equation}
where $\alpha$ and $\lambda$ are scale factors and $\mathcal{L}_{U}$ can be one or a combination of $\mathcal{L}_{E(i)}$, $\mathcal{L}_{D}$ and $\mathcal{L}_{HRL}$.

\section{Experiments}\label{sec:experiments}

\subsection{Data and pre-processing}\label{sec:experiments_data}
\noindent Four public medical image datasets are used in this work to evaluate the SR performance and robustness of the proposed method in simulating the clinical situation as widely as possible. Experiments are designed and applied on: the OASIS [\cite{OASIS}] dataset of single-modality brain MR scans; the BraTS [\cite{BraTSCite1, BraTSCite2, BraTSCite3}] dataset of multi-modal brain MR scans; the ACDC [\cite{ACDC}] dataset of cardiac MR images; and the COVID [\cite{dataset_COVID_for_RDST}] dataset of chest CT scans. Notice that experiments of ablation studies are mainly conducted with the OASIS dataset because it is clean and representative in the discussion of model architectures, hyper-parameters, loss functions and model efficiency. 
\\
\\
\textbf{OASIS} We randomly select 39 subjects (30 for training and 9 for testing) from the OASIS-brain dataset\footnote{OASIS: https://www.oasis-brains.org/} for the $\times 4$ super-resolution simulation experiments. Each subject includes 3 or 4 T1-weighted MRI scans and corresponding segmentation labels of one patient with early-stage Alzheimer's Disease (AD). Only one scan (T88-111) is used in this work, with an original size of $[176 \times 208 \times 176]$. It includes plenty of black background regions, providing useless information and slowing down the training process. Thus, the original scans and their corresponding segmentation labels are first rotated to the axial plane and centrally cropped to 145 slices of size $[160 \times 128]$. As a result, the OASIS training dataset includes 4350 slices, and the testing dataset includes 1305 images.
\\
\\
\textbf{BraTS} The BraTS dataset\footnote{BraTS: https://www.med.upenn.edu/cbica/brats2020/data.html} consists of multi-modal MRI scans of 285 patients, including 210 cases with glioblastoma and 75 cases with lower grade glioma. Scans of each patient include 4 registered MR modalities: native (T1), post-contrast T1-weighted (T1Gd), T2-weighted (T2) and T2 FLuid Attenuated Inversion Recovery (FLAIR). Manual annotations of the enhancing tumour (ET), the peritumoral edema (ED) and the necrotic and non-enhancing tumour core (NCR/NET) are also provided with each scan. In this work, we randomly select 120 patients from the BraTS dataset for training and 30 other patients for testing. In pre-processing, only slices with tumours are chosen for a fair comparison in the downstream segmentation task. As a result, there are 7333 slices for training and 1853 slices for testing. To remove the pure black background, all slices are centrally cropped to $[192 \times 192]$.
\\
\\
\textbf{ACDC}  The ACDC dataset\footnote{ACDC: https://www.creatis.insa-lyon.fr/Challenge/acdc/databases.html} includes 1.5T and 3.0T cardiac MR scans of 150 patients consisting of 5 evenly divided subgroups: normal subjects, previous myocardial infarction, dilated cardiomyopathy, hypertrophic cardiomyopathy and abnormal right ventricle. Additionally, the contours of the left ventricle (LV), right ventricle (RV) and myocardium are manually drawn and double-checked by two independent experts with more than 10 years of experience. These segmentation labels of 100 patients are released to the public. In this work, we randomly divide these 100 patients for training (80 patients with 1462 slices) and testing (20 patients with 373 slices). For a fair comparison, all slices are first centrally cropped to $[128 \times 128]$.
\\
\\
\textbf{COVID} The COVID dataset\footnote{COVID-19 CT: https://zenodo.org/record/3757476} includes 3D CT scans with left lung, right lung, and infection annotations of 20 COVID-19 patients. The proportion of infections in the lungs ranges from 0.01\% to 59\%. Annotations of the left lung, right lung and infection are manually labelled by experienced radiologists. In total, there are 300+ infections with 1800+ slices of various shapes. In this work, we uniformly crop a $[512 \times 512]$ region in the centre of each slice and randomly divide all scans to the training dataset (16 scans with 2264 slices) and the testing dataset (4 scans with 588 slices).
\\
\\
\textbf{HR-LR image pair generation} The original slices are used as high-resolution ground truth images $I_{hr} \in \mathbb{R}^{H\times W\times c}$, and the corresponding low-resolution images are generated by down-sampling:
\begin{equation}
    I_{lr} = (I_{hr}\otimes k)\downarrow_{s}+n, \forall I_{hr}\in \mathbb{R}^{H\times W\times c}
\end{equation}

\noindent where $k$ is a bicubic down-sampling kernel and $n$ is an additive Gaussian noise. we focus on $\times 4$ super-resolution tasks in this work, so the HR and LR patches are cropped with size $[96 \times 96]$ and $[24 \times 24]$, respectively.

\subsection{Evaluation Metrics}
\noindent In addition to Peak Signal-to-Noise Ration (PSNR) and Structural Similarity (SSIM), the SR results of the proposed RDST and SOTA methods are also evaluated in downstream segmentation tasks. As described in Section \ref{sec:mtd_unet}, we first train a segmentation U-Net for each dataset with HR images in the training subset and their corresponding ground truth segmentation labels. Take the OASIS dataset as an example. The pre-trained U-Net achieves high segmentation performance on HR images in the testing dataset, so the segmentation-based SR performance measurement can be reliable. To evaluate the SR results of SOTA methods and the proposed RDST variants, we use dice coefficients of the whole region and tissues depending on each dataset. For example, the experiments with the OASIS dataset involve the dice coefficient scores on the whole brain (Dice-T), grey matter (Dice-G), white matter (Dice-W) and cerebrospinal fluid (Dice-CSF):

\begin{align}
    &P_{sr} = U(I_{sr}), \nonumber \\
    &\text{Dice-T} = Dice(P_{sr}, L_{GT}), \nonumber \\
    &\text{Dice-G} = Dice(P_{sr}[C_{G}], L_{GT}[C_{G}]), \nonumber \\
    &\text{Dice-W} = Dice(P_{sr}[C_{W}], L_{GT}[C_{W}]), \nonumber \\
    &\text{Dice-CSF} = Dice(P_{sr}[C_{CSF}], L_{GT}[C_{CSF}]), \label{eqt:rdst_dice_score}
\end{align}
where $C_{G}$, $C_{W}$, and $C_{CSF}$ are label indexes of grey matter, white matter and CSF, respectively. Similarly, we use dice coefficient scores of the left ventricular cavity (Dice-LV), the right ventricular cavity (Dice-RV), the myocardium (Dice-MC) and the whole region (Dice-T) for the ACDC dataset and the dice coefficient scores of the left lung (Dice-LL), the right lung (Dice-RL), the lesion (Dice-Lesion) and the whole region (Dice-T) for the COVID dataset. In the experiments with the BraTS dataset, we use dice coefficient scores of the enhancing tumour (Dice-ET, including ET only), the tumour core (Dice-TC, including ET and NCR/NET) and the whole tumour (Dice-WT, including ET, ED and NCR/NET).

\subsection{Implementation details}\label{sec:rdst_implementation}
\noindent The proposed RDST and SOTA models are implemented with PyTorch [\cite{PyTorch}]. All experiments were performed on an Nvidia Quadro RTX 8000 GPU. Inspired by SwinIR [\cite{swinir}], the window size, attention head number and the basic feature embedding dimension are set to $[8 \times 8]$, 6 and 60, respectively. As mentioned in Section \ref{sec:method_rdstb}, each RDSTB module consists of 3 DSTBs with a growth rate of 30. Each DSTB module consists of 2 STLs. In this work, we propose two RDST variants. The original one consists of 8 RDSTBs, while the more efficient version (RDST-E) consists of only 4 RDSTBs. In the experiments, all parameters are initialised by Kaiming-uniform [\cite{HeInit}] and optimised by the Adam optimiser [\cite{Adam}]. we set the batch size to 32 for each step for both training stages. In the first training stage, the initial learning rate is set to 0.0002 with no decay, and the RDST is trained for 100k steps with only native $\mathcal{L}_{1}$ loss. The fine-tuning stage includes 20k steps. Its learning rate is initialised as 0.0001 and halved at $[10\text{k}, 15\text{k}, 17.5\text{k}]$. The scale factors in Equation \ref{eqt:rdst_loss_finetune} are set as $\alpha =1, \lambda = 10$ to ensure the segmentation-based perceptual loss dominates the fine-tuning stage. The L1 distance between feature maps of the first encoder block $\mathcal{L}_{E(1)}$ is mainly used as the perceptual loss, and other variations of $\mathcal{L}_{U}$ are used in ablation study.
\\
\\
\noindent The segmentation U-Net model is implemented with Segmentation-Models-PyTorch [\cite{SMP}]. It consists of 5 ResNet-based [\cite{ResNet}] encoder blocks and a decoder of native convolutional layers (Fig. \ref{fig:rdst_unet_segloss}). The channel number is set to 64 basically and doubled after each encoder block. This model is trained with Dice loss (Equation \ref{eqt:rdst_seg_loss_dice}) for 100k steps with the Adam optimiser. The learning rate is initialised as 0.0001 and halved at $[50\text{k}, 75\text{k}]$. 

\begin{figure*}[t]
    \centering
    \includegraphics[width=\textwidth]{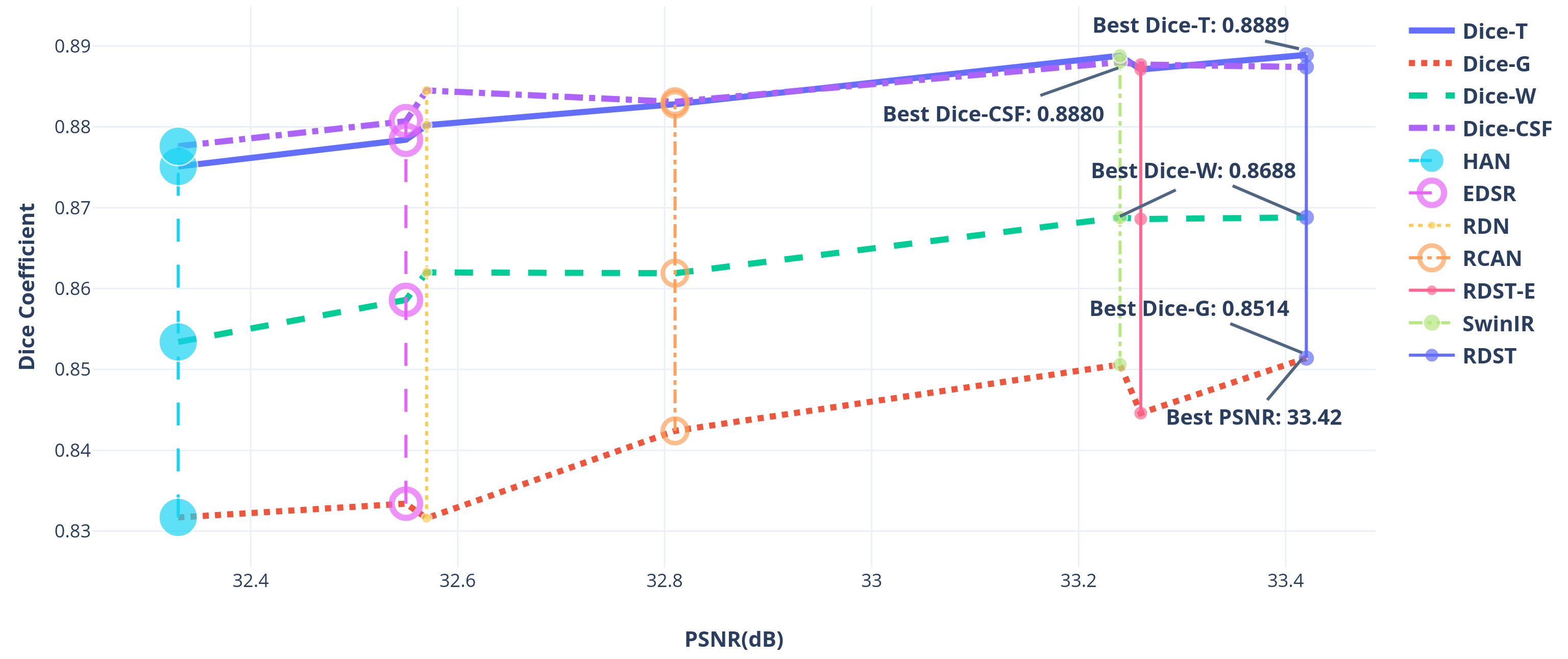}
    \caption{Comparing RDST and RDST-E with SOTA SISR methods on $\times 4$ super-resolution task with the OASIS dataset. The PNSR (the X-axis) indicates the SR image quality, while the dice scores (the Y-axis) represent the downstream segmentation performance. Dice-[T, G, W, CSF] specify the dice coefficient of whole brains, grey matter, white matter and cerebrospinal fluid, respectively. The best PSNR and dice scores are pointed out. Bigger symbols indicate more parameters of the models.}
    \label{fig:rdst_rst_sota}
\end{figure*}

\begin{figure*}[t]
    \centering
    \includegraphics[width=\textwidth]{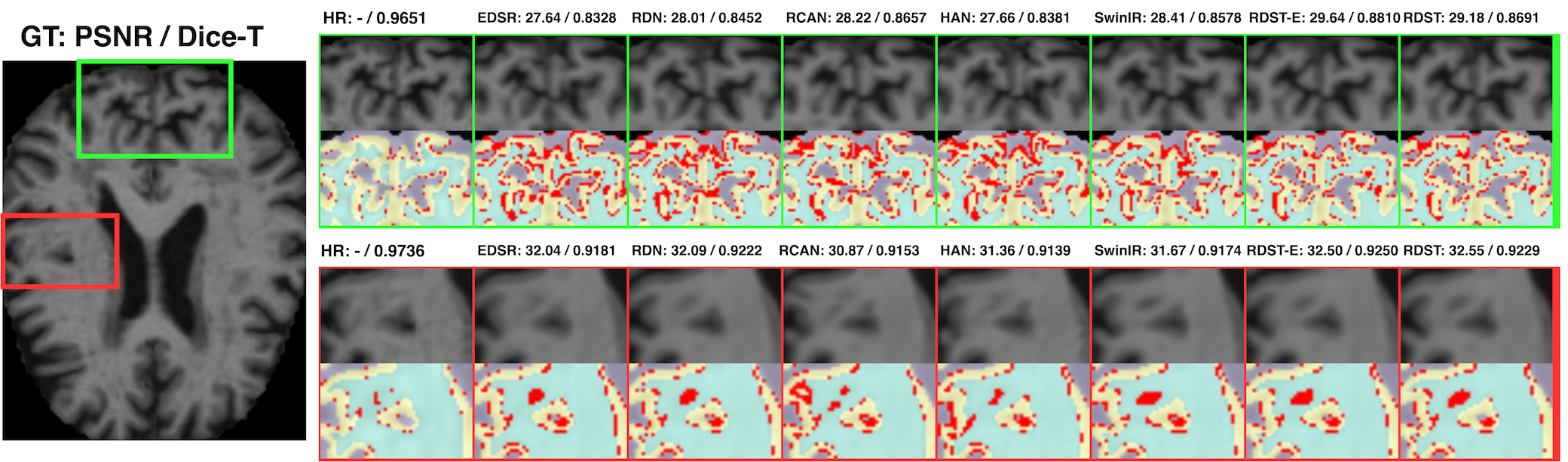}
    \caption{Comparing RDST with SOTA methods on $\times 4$ super-resolution task with OASIS dataset. SR results and corresponding segmentation predictions of a randomly selected slice are shown with PSNR and dice coefficient of the whole brain. In the segmentation labels, grey, yellow and cyan indicate grey matters, white matters and CSFs, respectively, and segmentation errors are marked as red.}
    \label{fig:rdst_oasis_sota}
\end{figure*}

\begin{table*}[h]
\caption{Compare RDST with SOTA methods on the OASIS dataset. The best and second-best scores are highlighted in red and blue respectively. MACs are calculated with an input size $[1 \times 40 \times 32 \times 1]$.}\label{tab:RDST-SOTA-Oasis}
\centering
\resizebox{\textwidth}{!}{
\begin{tabular}{l|llllll|rr}
\hline
\textbf{Mean(std)} & \textbf{PSNR(dB)$\uparrow$}                & \textbf{SSIM$\uparrow$}                       & \textbf{Dice-T$\uparrow$}                      & \textbf{Dice-G$\uparrow$}                     & \textbf{Dice-W$\uparrow$}                     & \textbf{Dice-CSF$\uparrow$}                   & \multicolumn{1}{l}{\textbf{MACs(G)$\downarrow$}} & \multicolumn{1}{l}{\textbf{params(M)$\downarrow$}} \\ \hline
\textbf{HR}        &                                            &                                               & 0.9520(0.012)                                  & 0.9401(0.040)                                 & 0.9399(0.013)                                 & 0.9408(0.015)                                 & \multicolumn{1}{l}{}                             & \multicolumn{1}{l}{}                               \\
\textbf{Bicubic}   & 29.88(2.7)                                 & 0.8574(0.052)                                 & 0.8125(0.0088)                                 & 0.7179(0.075)                                 & 0.8137(0.021)                                 & 0.8207(0.016)                                 & \multicolumn{1}{l}{}                             & \multicolumn{1}{l}{}                               \\ \hline
\textbf{EDSR}      & 32.55(3.5)                                 & 0.9184(0.039)                                 & 0.8784(0.0098)                                 & 0.8334(0.055)                                 & 0.8586(0.021)                                 & 0.8807(0.015)                                 & 64.22                                            & 43.08                                              \\
\textbf{RDN}       & 32.57(3.2)                                 & 0.9241(0.036)                                 & 0.8802(0.010)                                  & 0.8316(0.059)                                 & 0.8620(0.021)                                 & 0.8845(0.015)                                 & 7.95                                             & 5.76                                               \\
\textbf{RCAN}      & 32.81(3.6)                                 & 0.9224(0.038)                                 & 0.8828(0.0094)                                 & 0.8424(0.054)                                 & 0.8619(0.021)                                 & 0.8831(0.013)                                 & 41.34                                            & 32.03                                              \\
\textbf{HAN}       & 32.33(3.7)                                 & 0.9120(0.042)                                 & 0.8751(0.0091)                                 & 0.8317(0.055)                                 & 0.8534(0.022)                                 & 0.8776(0.015)                                 & 83.86                                            & 64.19                                              \\
\textbf{SwinIR}    & 33.24(3.7)                                 & 0.9287(0.036)                                 & {\color[HTML]{2972F4} \textbf{0.8888(0.0097)}} & {\color[HTML]{2972F4} \textbf{0.8506(0.054)}} & {\color[HTML]{E74025} \textbf{0.8688(0.020)}} & {\color[HTML]{E74025} \textbf{0.8880(0.015)}} & 14.68                                            & 11.47                                              \\
\textbf{RDST-E}    & {\color[HTML]{2972F4} \textbf{33.26(3.4)}} & {\color[HTML]{2972F4} \textbf{0.9291(0.035)}} & 0.8871(0.0096)                                 & 0.8446(0.057)                                 & {\color[HTML]{2972F4} \textbf{0.8686(0.020)}} & {\color[HTML]{2972F4} \textbf{0.8877(0.015)}} & {\color[HTML]{E74025} \textbf{3.53}}             & {\color[HTML]{E74025} \textbf{2.35}}               \\
\textbf{RDST}      & {\color[HTML]{E74025} \textbf{33.42(3.7)}} & {\color[HTML]{E74025} \textbf{0.9299(0.035)}} & {\color[HTML]{E74025} \textbf{0.8889(0.0097)}} & {\color[HTML]{E74025} \textbf{0.8514(0.054)}} & {\color[HTML]{E74025} \textbf{0.8688(0.021)}} & 0.8874(0.015)                                 & {\color[HTML]{2972F4} \textbf{6.17}}             & {\color[HTML]{2972F4} \textbf{4.40}}               \\ \hline
\end{tabular}
}
\end{table*}

\section{Results and discussion}\label{sec:result}
\noindent In this section, we first illustrate the superior performance of the proposed RDST variants compared with SOTA SISR methods on four medical image datasets, then discuss the key factors of RDST twofold: the novel architecture and the new segmentation-based perceptual loss.

\subsection{Comparing RDST with SOTA methods}\label{sec:result_sota}
\noindent Two variations of RDST are compared with 5 popular and representative state-of-the-art SISR methods, including: (1). pure convolutional methods EDSR [\cite{EDSR}] and RDN [\cite{RDN}]; (2) CNN based attention models RCAN [\cite{RCAN}] and HAN [\cite{HAN}]; and (3) self-attention-based vision transformers SwinIR [\cite{swinir}]. Additionally, we have done experiments with related SR methods in a wider range, such as zero-shot super-resolution [\cite{ZSSR}], scale-free super-resolution [\cite{MetaSR, MIASSR}] and pre-trained ViT [\cite{IPT}]. However, we finally decided not to involve these methods in the comparison because of their mediocre performance. 
\\
\\
\noindent RDST variants achieve the best image quality on all four datasets. Specifically, RDST-E achieves the best PSNR performance on the ACDC dataset, and RDST achieves the best PSNR scores on the OASIS dataset, the COVID dataset and every MR modality in the BraTS dataset. On average, RDST increases by 0.09dB in PSNR, and RDST-E leads to a 0.06dB increase compared with the most recent SOTA method SwinIR. Meanwhile, we clearly show that improving image quality can lead to significantly better performance in the downstream segmentation tasks. With the well-trained U-Net segmentation models and introducing the segmentation-based SR results evaluation, SOTA SISR methods considerably narrow the gap of segmentation accuracy between HR GT images and $\times 4$ bicubic interpolated images. In summary, RDST achieves the best dice coefficient scores of 8 targeted regions among all 15 regions. Detailed results of each dataset are as follows.
\\
\\
\noindent \textbf{Performance on the OASIS dataset} RDST achieves the best performance of almost all metrics in the $\times 4$ super-resolution experiment with the OASIS dataset (Fig. \ref{fig:rdst_rst_sota}). It brings noticeable improvements in image quality (+0.18dB PSNR and +0.0012 SSIM) to SwinIR. In the downstream segmentation task, SR results of RDST get the best dice coefficient scores of the whole brain, the grey matter and the white matter (Table \ref{tab:RDST-SOTA-Oasis}). The pre-trained U-Net achieves reliable segmentation performance on HR GT images. It clearly shows a notable decline with native bicubic interpolation SR results: $[0.1395, 0.2210, 0.1262, 0.1201]$ on whole brains, grey matters, white matters and CSFs, respectively. The proposed RDST narrows these gaps by $[0.0774, 0.1335, 0.0551, 0.0667]$ respectively. Notice that room for improvement exists as the best segmentation dice scores of all SR images are still significantly lower than HR images ($[-0.0621, -0.0887, -0.0711, -0.0518]$). On the other hand, the smallest model RDST-E achieves the second-best PSNR and SSIM scores with a slight decrease in segmentation performance. While visualising the SR results and their corresponding segmentation predictions, the segmentation labels can help determine the differences between SR images of SOTA methods, which are difficult to recognise with only the images. Vision transformers (i.e. SwinIR and RDST) achieve superior image quality on edges than CNNs (green box in Fig. \ref{fig:rdst_oasis_sota}). On the other hand, it is still challenging for all methods to reconstruct rich textures of small regions (red box in Fig. \ref{fig:rdst_oasis_sota}).
\\

\begin{figure*}
    \centering
    \includegraphics[width=\textwidth]{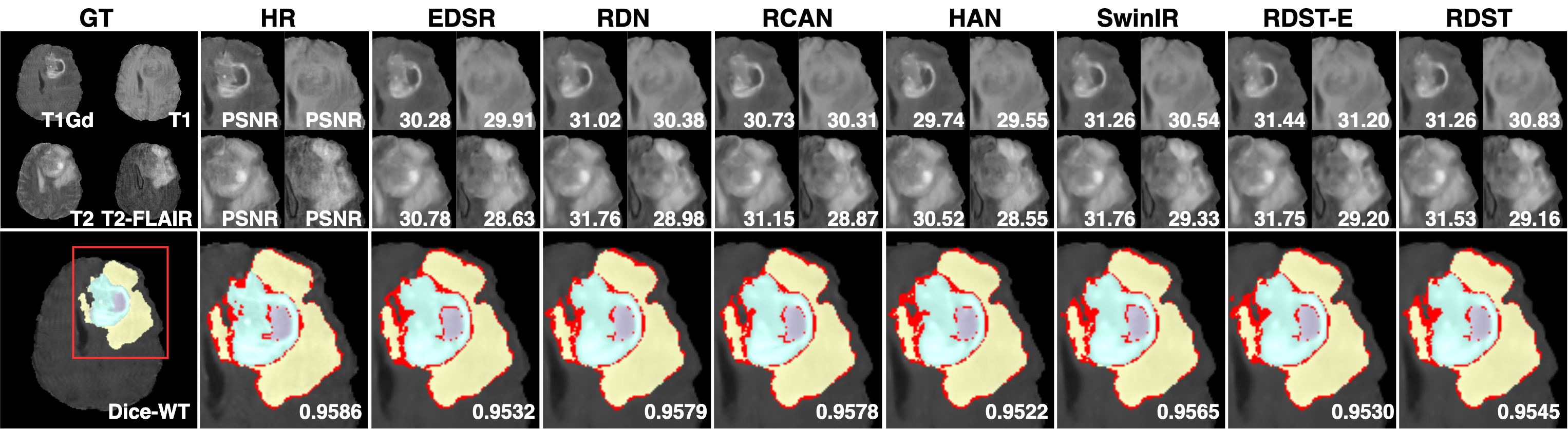}
    \caption{SR results of a random slice in the testing subset of BraTS. SR results of whole slices are generated, but only the regions within the tumour of the four modalities are plotted with predicted labels in the downstream segmentation task for better comparison. Annotations of tumour sub-regions are the necrotic and the non-enhancing (NCR \& NET) parts of the tumour in yellow, the peritumoral edema (ED) in cyan and the enhancing tumour in grey. Segmentation errors are indicated in red. PSNR and dice coefficient of the whole tumour (Dice-WT) are also displayed.}
    \label{fig:rdst_brats_sota}
\end{figure*}

\noindent \textbf{Model efficiency} Additionally, we compare the model efficiency in the experiment on the OASIS dataset by calculating the number of parameters and the Multi-Add Calculations (MACs) with an $[1 \times 40 \times 32 \times 1]$ input (Table \ref{tab:RDST-SOTA-Oasis}). RDST-E is the smallest and requires the fewest MACs, while RDST is the second smallest and requires fewer calculations than SOTA methods. Compared with SwinIR, RDST has only 38\% parameters, and RDST-E has only 20\% parameters. As a result, they reduce the computational cost by 58\% and 76\%, respectively.
\\
\\
\noindent \textbf{Performance on the BraTS dataset} Doing $\times 4$ super-resolution in the multi-modal brain tumour segmentation dataset is more challenging for the following reasons. First, the super-resolution method must be synchronously applied on four registered MR scans (i.e. T1Gd, T1, T2 and T2-FLAIR) so the input and output layers of RDST variants and SOTA methods are modified. Second, the downstream multi-modal tumour issue segmentation is challenging, and the U-Net with poor segmentation performance may misdirect the fine-tuning stage. In this experiment, the pre-trained U-Net has achieved dice coefficients of $[0.7830, 0.6919, 0.6820]$ on the whole tumour, the enhancing tumour and the tumour core, respectively. Compared with bicubic interpolation, SOTA deep neural networks significantly improve the SR performance (Table. \ref{tab:RDST-BraTS-SOTA}) on image quality and downstream segmentation performance. The proposed RDST achieves the best PSNR scores on all modalities, and the efficient version RDST-E achieves the second-best scores on three modalities (i.e. T1Gd, T1 and T2-FLAIR). SwinIR, the SOTA vision transformer for SISR tasks, also performs better than CNN models. It dominates the SSIM scores with RDST. On average, RDST gets +0.13dB higher PSNR than SwinIR and equal SSIM (-0.0003) of the four modalities. In the downstream tumour segmentation task, RDST achieves the best dice coefficients of the whole tumours and the enhancing tumours. EDSR achieves the best segmentation performance of the tumour cores. Interestingly, SR results of SwinIR and RDST can even provide more accurate segmentation labels than HR GT images, probably because the down-sample and super-resolve process removes some noises and misleading textures. Similar to the experiments on the OASIS dataset, the segmentation labels help to recognise the most challenging part in the SR image generation: reconstructing the blur edges and irregular textures (Fig. \ref{fig:rdst_brats_sota}). 
\\

\begin{figure*}
    \centering
    \includegraphics[width=\textwidth]{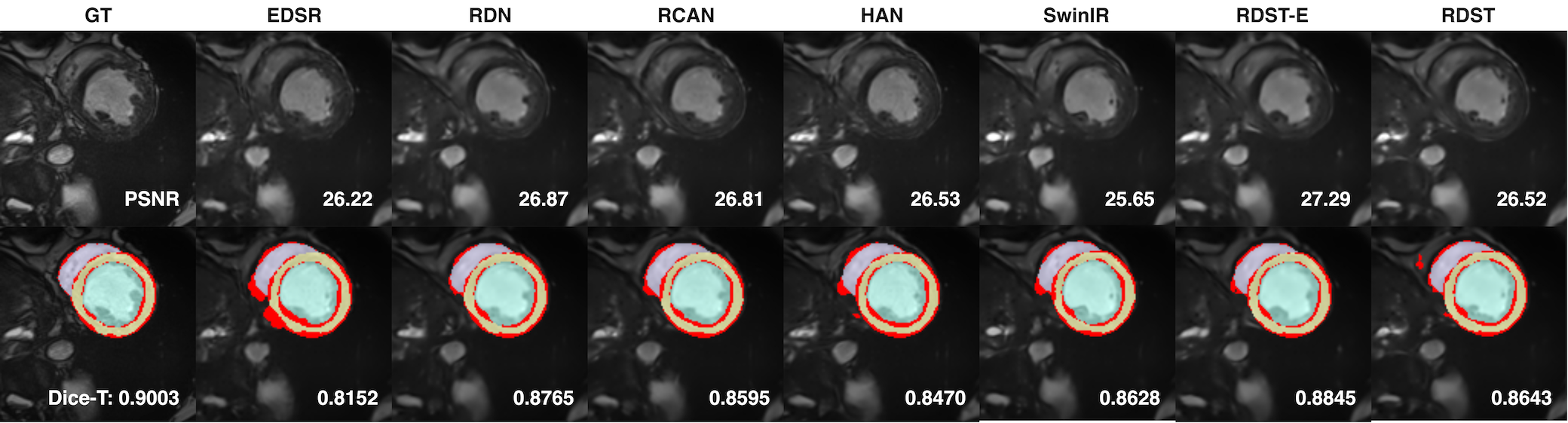}
    \caption{SR results of a random slice in the testing dataset of ACDC. Accurate segmentation predictions are annotated in grey (the RV cavity), yellow (the myocardium) and cyan (the LV cavity), while wrong predictions are annotated in red. PSNR scores and dice coefficients of the whole region are shown.}
    \label{fig:rdst_acdc_sota}
\end{figure*}

\begin{figure*}
    \centering
    \includegraphics[width=\textwidth]{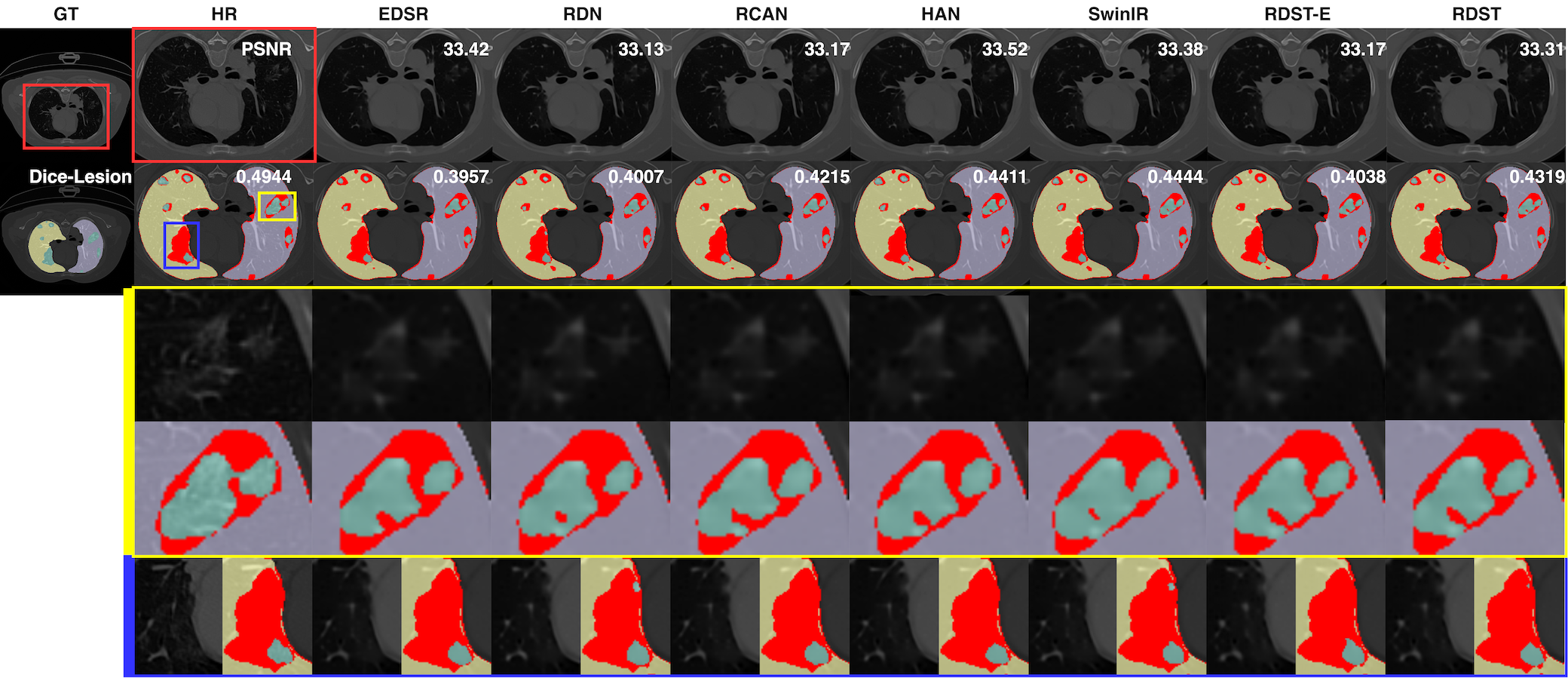}
    \caption{SR results of a random slice in the testing dataset of COVID-CT. PSNR scores and dice coefficients of the lesion are shown. Accurate segmentation predictions are annotated in grey (the left lung), yellow (the right lung) and cyan (the lesion), while wrong predictions are annotated in red. Patches of two sub-regions are zoomed in for better visualisation.}
    \label{fig:rdst_covid_sota}
\end{figure*}

\noindent \textbf{Performance on the ACDC dataset} Single image SR of cardiac MR images is challenging because the motion artefacts caused by patients' atrial fibrillation during the scanning procedure are individual and hard to resolve. As a result, data-driven deep learning methods can rarely bring a dramatic improvement in PSNR than traditional interpolation methods. In the comparison study of $\times 4$ magnification (Table. \ref{tab:RDST-ACDC-SOTA}), SOTA methods, including RDST variants, increase PSNR from +0.80 dB to +1.10 dB than bicubic interpolation. In contrast, SOTA SISR methods can easily lead to more than 3 dB improvement of PSNR on other datasets. Additionally, because the training data is limited (only 1462 slices with size $[128 \times 128]$), over-fitting may happen. As a result, the smallest model RDST-E achieves a significant advantage of PSNR (+0.20 dB higher than other methods). However, it is hard to evaluate the SR performance with only PSNR because most methods achieve very close scores from 27.03 dB to 27.06 dB. we guess the low PSNR scores of RDST and SwinIR are caused by the over-fitting of background noise, which leads to substantial pixel-wise errors but rare impacts in segmentation and visualisation (Fig. \ref{fig:rdst_acdc_sota}). In contrast, evaluating SR results with the dice coefficient scores and SSIM scores is more effective and robust. The segmentation U-Net is well-trained for the ACDC dataset with reliable performance on HR GT images. Meanwhile, the segmentation-based evaluation represents the global structure reconstruction accuracy in SR results. In this experiment, vision transformers (i.e. SwinIR, RDST-E and RDST) achieve the highest SSIM and perform the best in the downstream segmentation task, so we claim that they are better than the CNN-based methods.
\\
\\
\noindent \textbf{Performance on the COVID dataset} We also extend the proposed method to CT images. Compared with MR scans, CT scans are with higher resolution (e.g. $[512 \times 512]$) and fewer artefacts. All methods achieve very close PSNR (from 34.56 dB to 34.70 dB) and SSIM (from 0.8678 to 0.8707) scores in the $\times 4$ SR experiment (Table. \ref{tab:RDST-COVID-SOTA}). Specifically, RDST achieves the highest PSNR score and the best segmentation results of the whole region and the right lung. Meanwhile, RCAN achieves the best SSIM, and HAN achieves the best segmentation accuracy of the left lung and the lesion. Notice that both methods are with very deep architectures ($>800$ layers) and with more than 30M parameters, which may benefit the reconstruction of the textures in lesion areas (Fig. \ref{fig:rdst_covid_sota}). 

\begin{table*}[p]
\caption{Comparing RDST with SOTA methods in the $\times 4$ super-resolution task on the BraTS dataset. PSNR and SSIM scores are calculated on each modality (i.e. one of T1Gd, T1, T2, and T2-FLAIR) and averaged. Dice coefficients of the whole tumour (WT), the enhancing tumour (ET) and the tumour core (TC) are used in the downstream segmentation evaluation with the pre-trained 2D U-Net. The best and the second best scores are highlighted in red and blue, while $*$ indicates superior segmentation performance than HR GT images.}\label{tab:RDST-BraTS-SOTA}
\centering
\resizebox{\textwidth}{!}{
\begin{tabular}{l|lll|llll|llll}
\hline
\textbf{Mean(std)} & \multicolumn{3}{l|}{\textbf{Dice$\uparrow$}}                                                                                                & \multicolumn{4}{c|}{\textbf{PSNR$\uparrow$}}                                                                                                                                      & \multicolumn{4}{c}{\textbf{SSIM$\uparrow$}}                                                                                                                                                   \\
                   & \textbf{WT}                                   & \textbf{ET}                                  & \textbf{TC}                                  & \textbf{T1Gd}                              & \textbf{T1}                                & \textbf{T2}                                & \textbf{Flair}                             & \textbf{T1Gd}                                 & \textbf{T1}                                   & \textbf{T2}                                   & \textbf{Flair}                                \\ \hline
\textbf{HR}        & 0.7833(0.13)                                  & 0.6919(0.13)                                 & 0.6820(0.16)                                 & \textbf{}                                  & \textbf{}                                  & \textbf{}                                  & \textbf{}                                  &                                               &                                               &                                               &                                               \\
\textbf{Bicubic}   & 0.7614(0.12)                                  & 0.5226(0.20)                                 & 0.5878(0.18)                                 & 29.97(1.9)                                 & 29.04(2.3)                                 & 28.19(2.1)                                 & 29.33(2.4)                                 & 0.8571(0.035)                                 & 0.8661(0.037)                                 & 0.8509(0.037)                                 & 0.8362(0.047)                                 \\ \hline
\textbf{EDSR}      & 0.7800(0.12)                                  & 0.6854(0.12)                                 & {\color[HTML]{E74025} \textbf{0.6776(0.16)}} & 32.63(2.0)                                 & 32.36(2.4)                                 & 31.08(1.9)                                 & 31.89(2.4)                                 & 0.9143(0.024)                                 & 0.9271(0.025)                                 & 0.9156(0.024)                                 & 0.8968(0.033)                                 \\
\textbf{RDN}       & 0.7806(0.12)                                  & {\color[HTML]{2972F4} \textbf{0.6874(0.13)}} & {\color[HTML]{2972F4} \textbf{0.6729(0.17)}} & 32.97(2.0)                                 & 32.73(2.4)                                 & 31.48(2.0)                                 & 32.29(2.4)                                 & 0.9193(0.024)                                 & 0.9320(0.024)                                 & 0.9218(0.024)                                 & 0.9039(0.032)                                 \\
\textbf{RCAN}      & 0.7815(0.12)                                  & 0.6808(0.12)                                 & 0.6674(0.17)                                 & 32.84(2.0)                                 & 32.59(2.4)                                 & 31.29(2.0)                                 & 32.14(2.3)                                 & 0.9183(0.023)                                 & 0.9310(0.024)                                 & 0.9202(0.023)                                 & 0.9022(0.032)                                 \\
\textbf{HAN}       & 0.7801(0.12)                                  & 0.6833(0.12)                                 & 0.6722(0.17)                                 & 32.56(1.9)                                 & 32.24(2.4)                                 & 31.03(1.9)                                 & 31.84(2.3)                                 & 0.9155(0.024)                                 & 0.9281(0.024)                                 & 0.9168(0.024)                                 & 0.8976(0.033)                                 \\
\textbf{SwinIR}    & {\color[HTML]{E74025} \textbf{0.7836*(0.12)}} & 0.6816(0.12)                                 & 0.6710(0.16)                                 & 33.23(2.1)                                 & 33.03(2.5)                                 & {\color[HTML]{2972F4} \textbf{31.73(2.0)}} & 32.54(2.4)                                 & {\color[HTML]{2972F4} \textbf{0.9226(0.023)}} & {\color[HTML]{E74025} \textbf{0.9350(0.024)}} & {\color[HTML]{E74025} \textbf{0.9253(0.023)}} & {\color[HTML]{E74025} \textbf{0.9082(0.031)}} \\
\textbf{RDST-E}    & {\color[HTML]{2972F4} \textbf{0.7831(0.12)}}  & 0.6832(0.12)                                 & 0.6713(0.18)                                 & {\color[HTML]{2972F4} \textbf{33.32(2.0)}} & {\color[HTML]{2972F4} \textbf{33.13(2.4)}} & 31.71(2.1)                                 & {\color[HTML]{2972F4} \textbf{32.59(2.4)}} & 0.9218(0.024)                                 & {\color[HTML]{2972F4} \textbf{0.9339(0.025)}} & 0.9231(0.024)                                 & 0.9066(0.032)                                 \\
\textbf{RDST}      & {\color[HTML]{E74025} \textbf{0.7836*(0.12)}} & {\color[HTML]{E74025} \textbf{0.6883(0.12)}} & 0.6713(0.17)                                 & {\color[HTML]{E74025} \textbf{33.37(2.0)}} & {\color[HTML]{E74025} \textbf{33.22(2.5)}} & {\color[HTML]{E74025} \textbf{31.81(2.1)}} & {\color[HTML]{E74025} \textbf{32.63(2.4)}} & {\color[HTML]{E74025} \textbf{0.9227(0.023)}} & {\color[HTML]{E74025} \textbf{0.9350(0.024)}} & {\color[HTML]{2972F4} \textbf{0.9248(0.023)}} & {\color[HTML]{2972F4} \textbf{0.9075(0.032)}} \\ \hline
\end{tabular}
}
\end{table*}

\begin{table*}[p]
\caption{Comparing RDST with SOTA methods in the $\times 4$ super-resolution task on the ACDC dataset. In addition to PSNR and SSIM, dice coefficients of the total region (T), the left ventricular cavity (LV), the right ventricular cavity (RV), and the myocardium (MC) in the downstream segmentation task are used for evaluation. The best and the second-best scores are highlighted in red and blue respectively.}\label{tab:RDST-ACDC-SOTA}
\centering
\begin{tabular}{l|llllll}
\hline
\textbf{Mean(std)} & \textbf{PSNR$\uparrow$}                    & \textbf{SSIM$\uparrow$}                       & \textbf{Dice-T$\uparrow$}                     & \textbf{Dice-LV$\uparrow$}                    & \textbf{Dice-RV$\uparrow$}                   & \textbf{Dice-MC$\uparrow$}                    \\ \hline
\textbf{HR}        &                                            &                                               & 0.8932(0.027)                                 & 0.9184(0.034)                                 & 0.7390(0.11)                                 & 0.8783(0.036)                                 \\
\textbf{Bicubic}   & 26.14(2.7)                                 & 0.7501(0.053)                                 & 0.8096(0.051)                                 & 0.8717(0.059)                                 & 0.5927(0.15)                                 & 0.7697(0.058)                                 \\ \hline
\textbf{EDSR}      & 26.94(3.1)                                 & 0.7722(0.064)                                 & 0.8599(0.030)                                 & 0.8950(0.044)                                 & 0.6897(0.12)                                 & 0.8354(0.044)                                 \\
\textbf{RDN}       & {\color[HTML]{2972F4} \textbf{27.07(3.0)}} & 0.7854(0.058)                                 & 0.8624(0.029)                                 & {\color[HTML]{2972F4} \textbf{0.8979(0.038)}} & 0.6946(0.12)                                 & 0.8376(0.039)                                 \\
\textbf{RCAN}      & 27.06(3.0)                                 & 0.7819(0.061)                                 & 0.8657(0.030)                                 & 0.8947(0.044)                                 & 0.6867(0.14)                                 & 0.8399(0.042)                                 \\
\textbf{HAN}       & 27.06(3.4)                                 & 0.7738(0.069)                                 & 0.8666(0.030)                                 & 0.8946(0.043)                                 & 0.6971(0.13)                                 & 0.8419(0.046)                                 \\
\textbf{SwinIR}    & 27.04(3.1)                                 & {\color[HTML]{2972F4} \textbf{0.7876(0.059)}} & {\color[HTML]{E74025} \textbf{0.8705(0.026)}} & {\color[HTML]{E74025} \textbf{0.9001(0.043)}} & 0.6964(0.12)                                 & {\color[HTML]{E74025} \textbf{0.8456(0.039)}} \\
\textbf{RDST-E}    & {\color[HTML]{E74025} \textbf{27.24(3.0)}} & {\color[HTML]{E74025} \textbf{0.7940(0.057)}} & {\color[HTML]{2972F4} \textbf{0.8691(0.025)}} & 0.8954(0.043)                                 & {\color[HTML]{2972F4} \textbf{0.7008(0.12)}} & {\color[HTML]{2972F4} \textbf{0.8454(0.035)}} \\
\textbf{RDST}      & 27.03(3.1)                                 & 0.7875(0.059)                                 & {\color[HTML]{2972F4} \textbf{0.8691(0.027)}} & 0.8959(0.043)                                 & {\color[HTML]{E74025} \textbf{0.7071(0.11)}} & 0.8433(0.035)                                 \\ \hline
\end{tabular}
\end{table*}

\begin{table*}[p]
\caption{Comparing RDST with SOTA methods in the $\times 4$ super-resolution task on the COVID-CT dataset. In addition of PSNR and SSIM, dice coefficients of the total region (T), the left lung (LL), the right lung (RL), and the lesion in the downstream segmentation task are used for evaluation. The best and the second best scores are highlighted in red and blue respectively.}\label{tab:RDST-COVID-SOTA}
\centering
\begin{tabular}{l|llllll}
\hline
\textbf{Mean(std)} & \textbf{PSNR$\uparrow$}                    & \textbf{SSIM$\uparrow$}                      & \textbf{Dice-T$\uparrow$}                    & \textbf{Dice-LL$\uparrow$}                   & \textbf{Dice-RL$\uparrow$}                   & \textbf{Dice-Lesion$\uparrow$}               \\ \hline
\textbf{HR}        &                                            &                                              & 0.8762(0.11)                                 & 0.8553(0.17)                                 & 0.8905(0.10)                                 & 0.6554(0.035)                                \\
\textbf{Bicubic}   & 28.45(3.0)                                 & 0.7760(0.15)                                 & 0.8092(0.12)                                 & 0.7386(0.17)                                 & 0.8430(0.11)                                 & 0.3848(0.14)                                 \\ \hline
\textbf{EDSR}      & 34.59(4.4)                                 & 0.8694(0.15)                                 & 0.8315(0.18)                                 & 0.8265(0.20)                                 & {\color[HTML]{2972F4} \textbf{0.8713(0.13)}} & 0.5881(0.14)                                 \\
\textbf{RDN}       & 34.56(4.4)                                 & 0.8678(0.15)                                 & 0.8196(0.19)                                 & 0.8167(0.21)                                 & 0.8590(0.14)                                 & 0.5674(0.13)                                 \\
\textbf{RCAN}      & {\color[HTML]{2972F4} \textbf{34.69(4.4)}} & {\color[HTML]{E74025} \textbf{0.8707(0.15)}} & {\color[HTML]{2972F4} \textbf{0.8365(0.17)}} & 0.8175(0.21)                                 & 0.8684(0.13)                                 & {\color[HTML]{2972F4} \textbf{0.6207(0.11)}} \\
\textbf{HAN}       & 34.65(4.4)                                 & {\color[HTML]{2972F4} \textbf{0.8700(0.15)}} & 0.8323(0.18)                                 & {\color[HTML]{E74025} \textbf{0.8294(0.19)}} & 0.8695(0.13)                                 & {\color[HTML]{E74025} \textbf{0.6247(0.18)}} \\
\textbf{SwinIR}    & 34.66(4.5)                                 & 0.8698(0.15)                                 & 0.8299(0.18)                                 & 0.8201(0.20)                                 & 0.8678(0.13)                                 & 0.5846(0.12)                                 \\
\textbf{RDST-E}    & 34.62(4.5)                                 & 0.8678(0.15)                                 & 0.8264(0.19)                                 & 0.8134(0.21)                                 & 0.8620(0.15)                                 & 0.5709(0.088)                                \\
\textbf{RDST}      & {\color[HTML]{E74025} \textbf{34.70(4.5)}} & 0.8687(0.15)                                 & {\color[HTML]{E74025} \textbf{0.8445(0.16)}} & {\color[HTML]{2972F4} \textbf{0.8281(0.19)}} & {\color[HTML]{E74025} \textbf{0.8761(0.13)}} & 0.5884(0.085)                                \\ \hline
\end{tabular}
\end{table*}

\begin{figure*}[htp]
    \centering
    \includegraphics[width=0.9\textwidth]{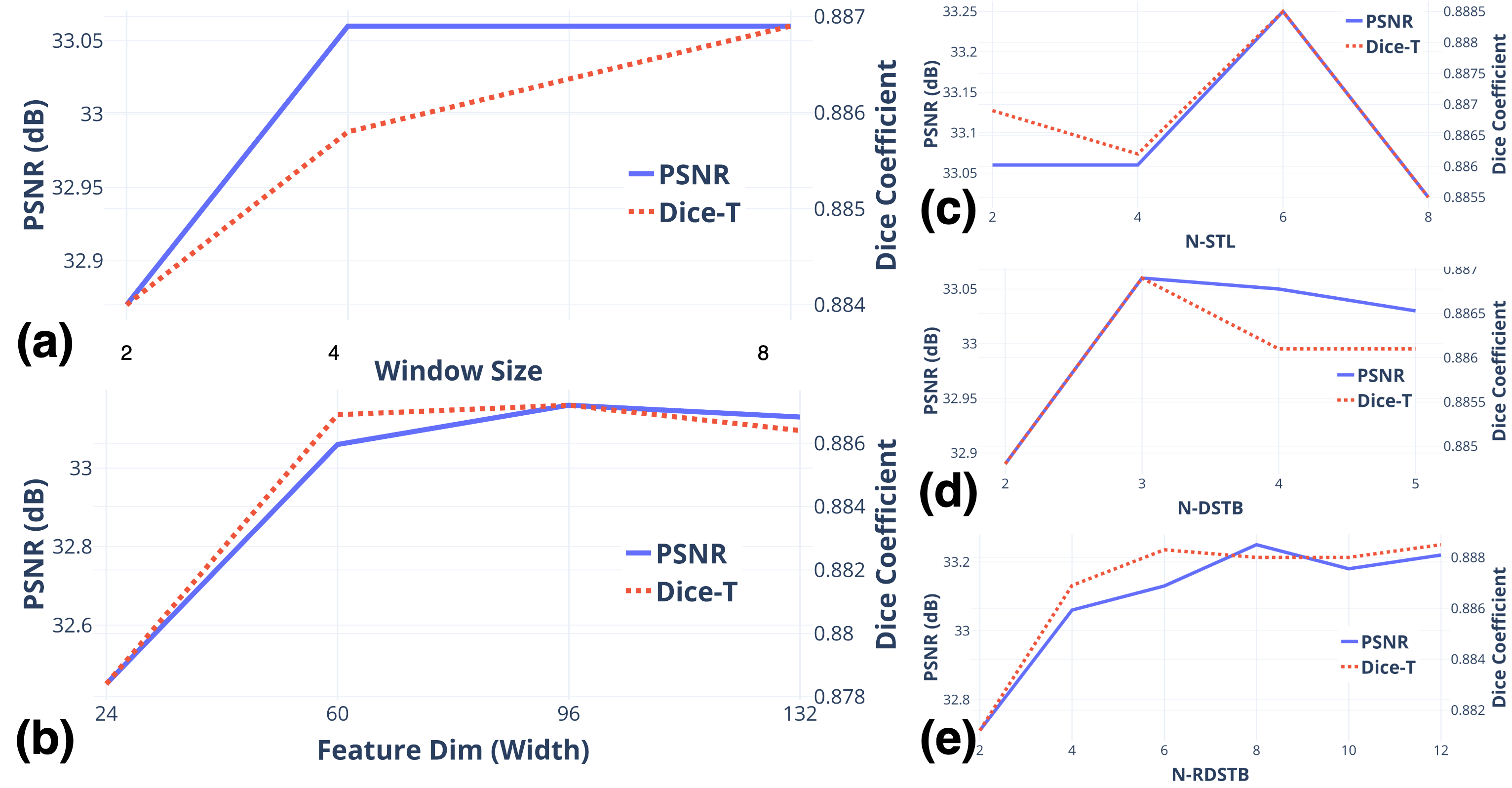}
    \caption{An ablation study on the hyperparameters of RDST: (a) impacts of the window size, varying in [2, 4, 8]; (b) impacts of the feature map embedding dimension, varying from 24 to 132; (c) impacts of the number of swin transformer layers (STLs) in the dense block, varying from 2 to 8; (d) impacts of the number of DSTBs in each RDSTB, varying from 2 to 5; (e) impacts of the number of RDSTBs in RDST, varying from 2 to 12.}
    \label{fig:rdst_params}
\end{figure*}

\subsection{Network architecture, attention and inference efficiency}\label{sec:result_efficiency}
\noindent An ablation study is designed to figure out the critical factors of RDST variants that achieve superior performance than SOTA SISR methods on the OASIS dataset. PSNR, SSIM and dice coefficient of the whole brain are used as SR image evaluation metrics. Meanwhile, the numbers of MACs and parameters and the throughout frame rate during inference are used to measure the model efficiency. All methods are trained with the same settings (100k steps with $\mathcal{L}_{1}$ only) to avoid the impacts of the segmentation-based perceptual loss and additional training steps.
\\
\\
\noindent \textbf{Network architecture} Comparing RDST variants and SOTA methods, the main differences in network architecture are fourfold. First, a model can be created with only convolutional layers or a hybrid of transformers (e.g. STL) and CNNs. Second, the window size, which presents the receptive field of each layer, can be different. Third, the model widths, which indicate the feature map dimensions, vary from 64 to 256. Fourth, one network can be shallow (e.g. EDSR-lite with only 32 layers) or deep (e.g. HAN with 1610 layers). Thus, we divide all methods into four groups: (1) native CNN SISR methods, including EDSR and RDN; (2) native CNN models with large window size, which are implemented based on ConvNet [\cite{liu2022convnet}]; (3) deep CNN models with more than 800 layers, i.e. RCAN and HAN; and (4) STL based vision transformers such as SwinIR and RDST. we propose the ConvNet-based SISR method to discuss the impacts of the receptive field [\cite{luo2016ReceptiveField}], which is considered the critical factor for the success of vision transformers [\cite{ding2022LargeKernel}]. Meanwhile, the lite versions of EDSR, SwinIR and ConvNet are also involved in comparing small SISR models with RDST-E.
\\
\\
\textbf{STL or CNN?} First, vision transformers show more dramatically improved SR image quality and segmentation accuracy than CNN methods. Notice that SwinIR and RDST variants are all based on the Swin transformer layer, which learns from the shared weights and localised operation of convolutional layers. Additionally, CNN layers are used at intervals of STL blocks which further ensures the training stability of these hybrid methods and leads to superior performance than pure CNN methods. Second, the larger window size failed to extend the success with transformers to CNN models. Both versions of the ConvNet-based methods perform worse than all the other methods. In deep networks, the receptive field depends on the window size in one layer and the number of layers. Thus, deeper networks with small window sizes can have an equal receptive field with shallow networks with large window sizes. For CNNs, the former works better probably because more non-linear activation is essential for feature extraction. Third, deeper models consistently achieve better results in each group with similar network architectures. Notice that both increases in width and depth lead to an increase in parameters and an efficiency decrease. In this comparison study on medical image SR tasks, increasing the number of blocks is more effective. For example, RDST has more layers and a smaller width than SwinIR, leading to fewer computational costs and parameters. It finally achieves equal PSNR (+0.01 dB) and SSIM (-0.0001) scores with only 38\% parameters of SwinIR. 
\\

\begin{table*}[p]
\caption{Ablation study on the OASIS dataset to answer why transformers perform better than CNNs. PSNR, downstream segmentation dice score of the whole brain (Dice-T) and inference frame rate (FPS) are used as evaluation metrics for both image quality and model efficiency. The best (in red) and second-best (in blue) scores are in bold. All methods are divided into four groups: (1). native CNN methods include EDSR and RDN; (2). the proposed ConvNet-based SR methods with the large receptive fields in CNN; (3). very deep CNNs with attention; and (4). methods with Swin transformer layers. Notice that the global feature fusion (GFF) in RDN and RDST variants is considered elementary hierarchical attention, while self-attention in transformers is considered a superset of both channel and spatial attention. Very wide (dimensions $\ge$ 128) and deep (layers $\ge$ 128) networks are in bold.}\label{tab:rdst-attention}
\centering
\resizebox{\textwidth}{!}{
\begin{tabular}{l|llrr|ccc|lll|rrr}
\hline
                                      & \multicolumn{4}{c|}{\textbf{Network Architecture}}                                                                             & \multicolumn{3}{c|}{\textbf{Attention}}                                                                           & \multicolumn{3}{c|}{\textbf{Performance}}                                                                                                   & \multicolumn{3}{c}{\textbf{Efficiency}}                                                                              \\
                                      & \textbf{Layer} & \textbf{Window}       & \multicolumn{1}{l}{\textbf{Width}}              & \multicolumn{1}{l|}{\textbf{Depth}} & \multicolumn{1}{l}{\textbf{Channel}} & \multicolumn{1}{l}{\textbf{Spatial}} & \multicolumn{1}{l|}{\textbf{Layer}} & \textbf{PSNR$\uparrow$}                    & \textbf{SSIM$\uparrow$}                       & \textbf{Dice-T$\uparrow$}                      & \textbf{MACs(G)$\downarrow$}         & \textbf{params(M)$\downarrow$}       & \textbf{FPS$\uparrow$}                 \\ \hline
\textbf{EDSR-lite}                    & CNN            & $3 \times 3$          & 64                                              & 32                                  & \xmark                                & \xmark                                & \xmark                               & 32.40(3.1)                                 & 0.9192(0.037)                                 & 0.8766(0.0099)                                 & 2.51                                 & {\color[HTML]{2972F4} \textbf{1.52}} & {\color[HTML]{E74025} \textbf{325.02}} \\
\textbf{EDSR}                         & CNN            & $3 \times 3$          & \textbf{256}                                    & 64                                  & \xmark                                & \xmark                                & \xmark                               & 32.55(3.5)                                 & 0.9184(0.039)                                 & 0.8784(0.0098)                                 & 64.22                                & 43.08                                & 88.00                                  \\
\textbf{RDN}                          & CNN            & $3 \times 3$          & \multicolumn{1}{l}{\textbf{64$\rightarrow$256}} & \textbf{142}                        & \xmark                                & \xmark                                & \cmark                               & 32.57(3.2)                                 & 0.9241(0.036)                                 & 0.8802(0.010)                                  & 7.95                                 & 5.76                                 & 70.89                                  \\ \hline
\textbf{ConvNet-lite}                 & CNN            & \textbf{$7 \times 7$} & 64                                              & 48                                  & \xmark                                & \xmark                                & \xmark                               & 31.92(2.9)                                 & 0.9111(0.039)                                 & 0.8669(0.010)                                  & {\color[HTML]{2972F4} \textbf{1.69}} & {\color[HTML]{E74025} \textbf{0.88}} & {\color[HTML]{326FBA} \textbf{213.54}} \\
\textbf{ConvNet-large}                & CNN            & \textbf{$7 \times 7$} & \textbf{192}                                    & 96                                  & \xmark                                & \xmark                                & \xmark                               & 32.14(3.3)                                 & 0.9130(0.040)                                 & 0.8733(0.010)                                  & 21.00                                & 12.43                                & 98.79                                  \\ \hline
\textbf{RCAN}                         & CNN            & $3 \times 3$          & 64                                              & \textbf{1610}                       & \cmark                                & \xmark                                & \xmark                               & 32.81(3.6)                                 & 0.9224(0.038)                                 & 0.8828(0.0094)                                 & 41.34                                & 32.03                                & 6.38                                   \\
\textbf{HAN}                          & CNN            & $3 \times 3$          & \textbf{128}                                    & \textbf{811}                        & \cmark                                & \cmark                                & \cmark                               & 32.33(3.7)                                 & 0.9120(0.042)                                 & 0.8751(0.0091)                                 & 83.86                                & 64.19                                & 17.01                                  \\ \hline
\textbf{SwinIR-lite}                  & STL+CNN        & \textbf{$8 \times 8$} & 60                                              & 101                                 & \cmark                                & \cmark                                & \xmark                               & 32.87(3.2)                                 & 0.9252(0.037)                                 & 0.8836(0.0095)                                 & {\color[HTML]{E74025} \textbf{1.15}} & {\color[HTML]{E74025} \textbf{0.88}} & 30.12                                  \\
\textbf{SwinIR}                       & STL+CNN        & \textbf{$8 \times 8$} & \textbf{180}                                    & \textbf{150}                        & \cmark                                & \cmark                                & \xmark                               & {\color[HTML]{2972F4} \textbf{33.24(3.7)}} & {\color[HTML]{2972F4} \textbf{0.9287(0.036)}} & {\color[HTML]{E74025} \textbf{0.8888(0.0097)}} & 14.68                                & 11.47                                & 19.45                                  \\
\textbf{RDST-E($\mathcal{L}_{1}$)}   & STL+CNN        & \textbf{$8 \times 8$} & \multicolumn{1}{l}{\textbf{60$\rightarrow$150}} & 114                                 & \cmark                                & \cmark                                & \xmark                               & 33.06(3.3)                                 & 0.9278(0.035)                                 & 0.8869(0.0091)                                 & 3.53                                 & 2.35                                 & 28.47                                  \\
\textbf{RDST($\mathcal{L}_{1}$)}     & STL+CNN        & \textbf{$8 \times 8$} & \multicolumn{1}{l}{\textbf{60$\rightarrow$150}} & \textbf{226}                        & \cmark                                & \cmark                                & \xmark                               & {\color[HTML]{E74025} \textbf{33.25(3.5)}} & 0.9286(0.035)                                 & 0.8880(0.0097)                                 & 6.17                                 & 4.40                                 & 13.93                                  \\
\textbf{RDST+GFF($\mathcal{L}_{1}$)} & STL+CNN        & \textbf{$8 \times 8$} & \multicolumn{1}{l}{\textbf{60$\rightarrow$150}} & \textbf{228}                        & \cmark                                & \cmark                                & \cmark                               & 33.23(3.5)                                 & {\color[HTML]{E74025} \textbf{0.9290(0.035)}} & {\color[HTML]{2972F4} \textbf{0.8887(0.0095)}} & \cellcolor[HTML]{E5F6FF}6.17         & \cellcolor[HTML]{E5F6FF}4.40         & \cellcolor[HTML]{E5F6FF}13.89          \\ \hline
\end{tabular}
}
\end{table*}

\begin{figure*}[p]
    \centering
    \includegraphics[width=0.8\textwidth]{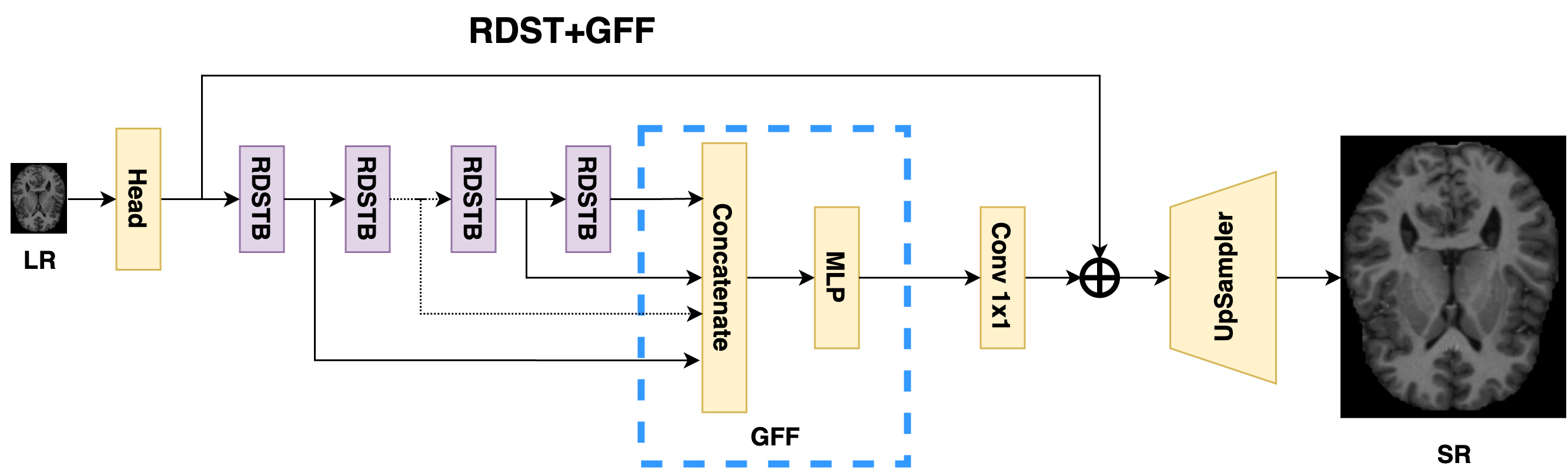}
    \caption{A RDST variant with MLP-based global feature fusion (GFF).}
    \label{fig:rdst_rst_rdstn}
\end{figure*}

\begin{figure*}[p]
    \centering
    \includegraphics[width=0.9\textwidth]{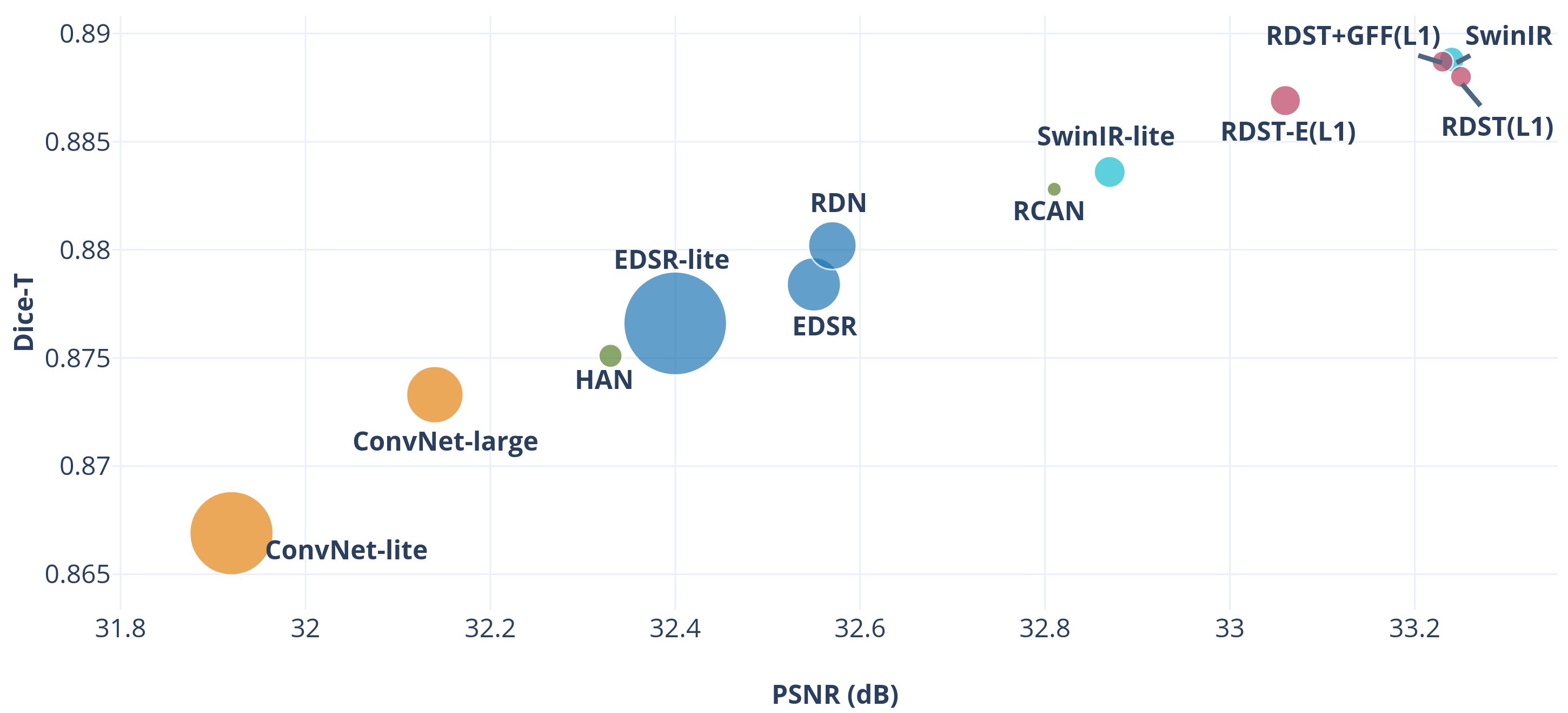}
    \caption{Ablation study on model architectures and attention mechanisms. All models are divided into 5 groups: native CNNs in blue, CNNs with a large receptive field in yellow, CNN-based attention models in green, SwinIR variations in light blue and the proposed RDST variations in red. PSNR scores (the x-axis) and dice coefficients of the whole brain in the downstream segmentation task (Dice-T, the y-axis) are used as evaluation metrics for SR image quality. Bigger symbols indicate higher inference throughout frame rates. For a fair comparison, all models apply the same training settings ($\mathcal{L}_{1}$ only for 100k steps with no extra fine-tuning).}
    \label{fig:rdst_rst_attention}
\end{figure*}

\noindent \textbf{Hyperparameters} Additionally, experiments are designed on RDST-E to figure out the impacts of hyperparameters of RDST (Fig. \ref{fig:rdst_params}). In RDST there is a gradual growth of feature map dimension because of the dense connections. we take the dimension after local feature fusion as the model width. On the other hand, a deeper RDST can be created by increasing three factors: the number of Swin transformer layers in the DSTB modules, the number of DSTBs and the number of RDSTBs. Briefly speaking, increasing the number of RDSTB modules, the window size and the model width in an adequate range can improve SR image quality.
\\
\\
\noindent \textbf{Attention} The attention mechanism is briefly regarded as threefold depending on where it works: channel attention, spatial attention and layer attention (Table. \ref{tab:rdst-attention}). Notice that the global feature fusion in RDN [\cite{RDN}] is considered elementary layer attention, and the self-attention in transformers is considered a super-set of both channel and spatial attention. Attention in CNN models raises training stability and leads to very deep networks but rarely improves the medial image super-resolution performance. For example, RCAN has 811 layers, and HAN has 1610 layers, but neither performs superior to other methods. On the other hand, the self-attention in vision transformers, which introduce dynamic parameters to the whole computation process, dramatically improves the SR performance as in SwinIR [\cite{swinir}] and RDST variations. Meanwhile, the GFF module has no noticeable improvement or decline in SR results, so it is abandoned in the final design of RDST.
\\
\\
\textbf{Inference frame rate} In addition to the number of parameters and calculations (MACs), we introduce the inference frame rate (FPS) as a more straightforward evaluation of model efficiency. Compared with MR and CT scanning in the clinic, all methods can almost real-time (at least 6 slices per second on an Nvidia RTX 8000 GPU for $[40 \times 32 \times 1] \rightarrow [160 \times 128 \times 1]$ SR. In addition to MACs and parameters, the inference efficiency depends on GPU acceleration and model architecture. Thus, transformers are generally slower than CNNs, although they have fewer parameters. Additionally, the depth of each model plays a crucial role in inference efficiency because it cannot be executed in parallel. For example, RCAN is the slowest model as it has the most layers. Among the vision transformers, RDST has fewer parameters and requires less computation but is slower than SwinIR because of more layers. The main advantage of vision transformers is that self-attention activates the capability of parameters more adequately, so superior performance is achieved with less computation and shallower networks. With potential hardware acceleration support in the future, transformers may be smaller and faster than CNNs. Specifically, RDST-E has a similar size to the lite version model (e.g. EDSR-lite and SwinIR-lite) but achieves comparable performance with regular SISR methods such as SwinIR.

\subsection{Impacts of the segmentation-based perceptual loss}\label{sec:result_loss}
\noindent In this section, we discuss the impacts of the proposed segmentation-based perceptual loss $\mathcal{L}_{U}$ with the following questions:

\begin{enumerate}
    \item Is the segmentation-based perceptual loss better than existing perceptual and adversarial losses? 
    \item Feature maps of which layer in the pre-trained U-Net should be used? 
    \item What connection has been created with this perceptual loss between super-resolution and segmentation tasks? 
    \item Can this perceptual loss improve SR performance with other SISR methods? 
    \item How can this segmentation-based perceptual loss be extended to datasets without segmentation labels?  
\end{enumerate}

\noindent To answer these questions (Table. \ref{tab:rdst-seg-loss-mode}), an RDST-E model is trained for 100k steps with only $\mathcal{L}_{1}$ as the \textbf{Baseline} and then fine-tuned for extra 20k steps with different combinations of $\mathcal{L}_{1}$ and perceptual losses. To avoid the impacts of the additional training steps, we further train an RDST-E with $\mathcal{L}_{1}$ for the same steps (so-called "\textbf{$\mathcal{L}_{1}$ only}").
\\
\\
\textbf{Comparison with L1, VGG and WGANGP} The shallow feature map based perceptual loss $\mathcal{L}_{E(1)}$ achieves the best SR image quality in all models. It results in significant increases of PSNR (+0.20 dB) and SSIM (+0.0013) than the Baseline (100k-steps training with only $\mathcal{L}_{1}$) and +0.14 dB PSNR and +0.0005 SSIM comparing with the 120k-steps $\mathcal{L}_{1}$ model, so the improvement is mainly caused by the proposed loss function but not the extra training steps. In contrast, neither the VGG-based perceptual loss nor the WGANGP loss combination can lead to better image quality. Actually, in the experiments, we have tested a big range ($0.0001< \gamma < +\infty$) of the scale factor of the VGG-based perceptual loss when using it with the $\mathcal{L}_{1}$ loss ($\mathcal{L}_{1}+\gamma \mathcal{L}_{VGG}$) and find that all the combinations decline the image quality in the medical image SR task. 
\\

\begin{table*}[pt]
\caption{Ablation study of the segmentation-based perceptual loss variations. $\mathcal{L}_{E(i)}$ indicates the native L1 distance between the feature maps of the $i$-th encoder block, while $\sum_{1}^{5}\mathcal{L}_{E(i)}$ indicates the sum. L1 distance between the outputs of the decoder ($\mathcal{L}_{D}$) and the dice loss of predicted labels ($\mathcal{L}_{HRL}$) are also tested. The best (in red) and the second best (in blue) scores are in bold.}
\label{tab:rdst-seg-loss-mode}
\centering
\begin{tabular}{l|llr|llll}
\hline
\textbf{Mean(std)}                        & \textbf{PSNR$\uparrow$}                    & \textbf{SSIM$\uparrow$}                       & \textbf{FID$\downarrow$}              & \textbf{Dice-T$\uparrow$}                      & \textbf{Dice-G$\uparrow$}                     & \textbf{Dice-W$\uparrow$}                     & \textbf{Dice-CSF$\uparrow$}                   \\ \hline
\textbf{Baseline}                         & 33.06(3.3)                                 & 0.9278(0.035)                                 & 81.63                                 & 0.8869(0.0091)                                 & 0.8438(0.057)                                 & 0.8685(0.020)                                 & 0.8877(0.014)                                 \\
\textbf{$\mathcal{L}_{1}$ only}                & {\color[HTML]{2972F4} \textbf{33.12(3.4)}} & {\color[HTML]{2972F4} \textbf{0.9286(0.035)}} & 81.30                                 & 0.8874(0.0094)                                 & 0.8453(0.057)                                 & 0.8688(0.020)                                 & 0.8880(0.014)                                 \\
\textbf{WGANGP+VGG}                       & 33.06(3.4)                                 & 0.9259(0.037)                                 & {\color[HTML]{2972F4} \textbf{72.64}} & 0.8885(0.0094)                                 & 0.8480(0.055)                                 & {\color[HTML]{2972F4} \textbf{0.8692(0.020)}} & 0.8883(0.014)                                 \\
\textbf{VGG only}                         & {\color[HTML]{2972F4} \textbf{33.12(3.4)}} & 0.9277(0.036)                                 & {\color[HTML]{E74025} \textbf{70.83}} & 0.8880(0.0092)                                 & 0.8471(0.056)                                 & 0.8689(0.020)                                 & 0.8882(0.014)                                 \\ \hline
\textbf{$\mathcal{L}_{E(1)}$}             & {\color[HTML]{E74025} \textbf{33.26(3.4)}} & {\color[HTML]{E74025} \textbf{0.9291(0.035)}} & 82.13                                 & 0.8871(0.0096)                                 & 0.8446(0.057)                                 & 0.8686(0.020)                                 & 0.8877(0.015)                                 \\
\textbf{$\mathcal{L}_{E(2)}$}             & 32.56(3.3)                                 & 0.9165(0.040)                                 & 73.07                                 & 0.8858(0.0089)                                 & 0.8484(0.052)                                 & 0.8650(0.020)                                 & 0.8854(0.015)                                 \\
\textbf{$\mathcal{L}_{E(3)}$}             & 32.53(3.3)                                 & 0.9181(0.040)                                 & 74.83                                 & 0.8855(0.0087)                                 & 0.8483(0.051)                                 & 0.8643(0.021)                                 & 0.8843(0.014)                                 \\
\textbf{$\mathcal{L}_{E(4)}$}             & 32.31(3.2)                                 & 0.9138(0.041)                                 & 82.77                                 & 0.8845(0.0087)                                 & 0.8473(0.051)                                 & 0.8630(0.021)                                 & 0.8849(0.014)                                 \\
\textbf{$\mathcal{L}_{E(5)}$}             & 32.04(3.2)                                 & 0.9061(0.041)                                 & 87.01                                 & 0.8830(0.0079)                                 & 0.8473(0.049)                                 & 0.8609(0.021)                                 & 0.8821(0.013)                                 \\
\textbf{$\mathcal{L}_{\sum_{1}^{5}E(i)}$} & 32.30(3.3)                                 & 0.9144(0.040)                                 & 77.02                                 & 0.8833(0.0085)                                 & 0.8469(0.050)                                 & 0.8612(0.021)                                 & 0.8821(0.014)                                 \\ \hline
\textbf{$\mathcal{L}_{D}$}                & 32.28(3.3)                                 & 0.8976(0.038)                                 & 95.91                                 & {\color[HTML]{2972F4} \textbf{0.8893(0.0086)}} & {\color[HTML]{2972F4} \textbf{0.8522(0.050)}} & 0.8687(0.021)                                 & {\color[HTML]{E74025} \textbf{0.8897(0.012)}} \\
\textbf{$\mathcal{L}_{HRL}$}              & 32.78(3.3)                                 & 0.9230(0.037)                                 & 78.02                                 & {\color[HTML]{E74025} \textbf{0.8899(0.0082)}} & {\color[HTML]{E74025} \textbf{0.8532(0.050)}} & {\color[HTML]{E74025} \textbf{0.8693(0.021)}} & {\color[HTML]{2972F4} \textbf{0.8890(0.013)}} \\ \hline
\end{tabular}
\end{table*}

\noindent \textbf{Comparison of $\mathcal{L}_{U}$ variations} As mentioned in Section \ref{sec:mtd_unet}, variants of the segmentation-based perceptual loss are defined depending on the choice of feature maps in the U-Net. Generally, researchers agree that shallow layers represent local and basic features while deep layers represent global and semantic information in networks. we conduct experiments to compare the $\mathcal{L}_{U}$ variations. Narrowing the distance between the feature maps of the first encoder block (i.e. $\mathcal{L}_{E(1)}$) in the segmentation U-Net is an effective restriction of pixel-wise and structure reconstruction, leading to an increase of PSNR and SSIM scores. On the other hand, decreasing the dice coefficient of the predicted labels (i.e. $\mathcal{L}_{HRL}$) and the distance between the output of decoders (i.e. $\mathcal{L}_{D}$) help semantic information recovery, leading to better segmentation performance. For example, the model fine-tuned with $\mathcal{L}_{HRL}$ achieves the best dice scores of the whole brain, the grey matter and the white matter and the second best dice score of the CSF with a slight decline of PSNR and SSIM. Meanwhile, the experiments show that outputs of both ends in the U-Net are more useful for the SR task, but the feature maps of hidden layers seem useless. Perceptual losses based on the feature maps of the second to the fifth encoder block neither increase the PSNR and SSIM scores nor improve the segmentation performance. 
\\
\\
\textbf{SR for humans or machines?} In addition to the distortion-perception trade-off of SR results, we find a human and machine perceptual difference in the experiments of perceptual losses. we admit that both PSNR and SSIM are essential for SR image evaluation of fidelity. However, neither can represent human perception or the performance in potential downstream image analysis tasks. Similarly, Fr\'echet Inception distance (FID [\cite{FID}]) has been used in previous works of medical image analysis tasks [\cite{yi2019GAN_for_MI_Review_MIA, kazeminia2020GAN_for_MIA_Review}] to evaluate the perceptual quality. However, its significance for medical images is also doubtful because the metric is designed for and pre-trained with natural images. In clinics, medical images are mainly for doctors to view and for machines to auto analysis. However, it is hard to explain how PSNR, SSIM and FID represent the perceptual performance in both cases. Inspired by [\cite{xia2021CMRSR}], we take segmentation as a typical downstream task to discuss the difference between human perception (with FID) and machine perception (with dice coefficients) of super-resolved medical images. Based on the conclusions and discussions in previous works [\cite{MIASSR}], we agree that PSNR and SSIM represent the fidelity of SR images and assume that FID denotes human perception quality. Additionally, we take the segmentation dice coefficient scores as the measurement for machine perception. In general, PSNR and SSIM closely correspond, but they are independent with either FID or dice scores. There is a trade-off between these three aspects, and none of the models can achieve superior performance in more than two directions. Thus, we suggest SR models be customised to suit the particular task. For example, the RDST variant fine-tuned with $\mathcal{L}_{E(1)}$ is proper for general purpose, and the variations fine-tuned with $\mathcal{L}_{VGG}$ and $\mathcal{L}_{WGANGP}$ are recommended for human viewing. Furthermore, we suggest using the segmentation label-based loss $\mathcal{L}_{HRL}$ for SR model fine-tuning to meet the particular needs of downstream segmentation tasks.
\\

\begin{table*}[t]
\caption{Fine-tuning RDST variations with $\mathcal{L}_{HRL}$ and comparing with SOTA methods in the downstream segmentation tasks. The mean and standard deviations of the dice coefficients of the whole organs (e.g. brain or tumour) of SR results are compared. The best and the second best scores are highlighted in red and blue, respectively, while $*$ indicates better segmentation performance than HR ground truth images. Red and green cells indicate improved and declined scores compared to $\mathcal{L}_{E(1)}$, respectively.}
\label{tab:rdst-hrl}
\centering
\begin{tabular}{l|llll|ll}
\hline
\textbf{Dice-T/WT$\uparrow$} & \textbf{HR}   & \textbf{Bicubic} & \textbf{EDSR}   & \textbf{RCAN}                                 & \textbf{RDST-E($\mathcal{L}_{E(1)}$)}                                  & \textbf{RDST($\mathcal{L}_{E(1)}$)}                                    \\ \hline 
\textbf{OASIS}               & 0.9520(0.012) & 0.8125(0.0088)   & 0.8784(0.0098) & 0.8828(0.0094)                                & 0.8871(0.0096)                                                         & 0.8889(0.0097)                                                         \\
\textbf{BraTS}               & 0.7833(0.13)  & 0.7614(0.12)     & 0.7800(0.12)   & 0.7815(0.12)                                  & 0.7831(0.12)                                                           & {\color[HTML]{2972F4} \textbf{0.7836*(0.12)}}                          \\
\textbf{ACDC}                & 0.8932(0.027) & 0.8096(0.051)    & 0.8599(0.030)  & 0.8657(0.030)                                 & 0.8691(0.025)                                                          & 0.8691(0.027)                                                          \\
\textbf{COVID-CT}            & 0.8762(0.11)  & 0.8092(0.12)     & 0.8315(0.18)   & 0.8365(0.17)                                  & 0.8264(0.19)                                                           & {\color[HTML]{E74025} \textbf{0.8445(0.16)}}                           \\ \hline \hline
\textbf{Dice-T/WT$\uparrow$} &               & \textbf{RDN}     & \textbf{HAN}   & \textbf{SwinIR}                              & \textbf{RDST-E($\mathcal{L}_{HRL}$)}                                   & \textbf{RDST($\mathcal{L}_{HRL}$)}                                     \\ \hline 
\textbf{OASIS}               &               & 0.8802(0.010)    & 0.8751(0.0091) & 0.8888(0.0097)                                & \cellcolor[HTML]{FFE9E8}{\color[HTML]{2972F4} \textbf{0.8899(0.0082)}} & \cellcolor[HTML]{FFE9E8}{\color[HTML]{E74025} \textbf{0.8906(0.0081)}} \\
\textbf{BraTS}               &               & 0.7806(0.12)     & 0.7801(0.12)   & {\color[HTML]{2972F4} \textbf{0.7836*(0.12)}} & \cellcolor[HTML]{EAFAF1}0.7825(0.13)                                   & \cellcolor[HTML]{FFE9E8}{\color[HTML]{E74025} \textbf{0.7842*(0.13)}}  \\
\textbf{ACDC}                &               & 0.8624(0.029)    & 0.8666(0.030)  & 0.8705(0.026)                                 & \cellcolor[HTML]{FFE9E8}{\color[HTML]{E74025} \textbf{0.8745(0.024)}}  & \cellcolor[HTML]{FFE9E8}{\color[HTML]{2972F4} \textbf{0.8732(0.024)}}  \\
\textbf{COVID-CT}            &               & 0.8196(0.19)     & 0.8323(0.18)   & 0.8299(0.18)                                  & \cellcolor[HTML]{FFE9E8}0.8347(0.18)                                   & \cellcolor[HTML]{EAFAF1}{\color[HTML]{2972F4} \textbf{0.8411(0.16)}}   \\ \hline
\end{tabular}
\end{table*}

\begin{table}[t]
    \caption{Dice coefficients of each tissue in the downstream segmentation tasks of SR results. RDST variations (fine-tuned with $\mathcal{L}_{HRL}$) and one SOTA method with the best performance are compared for each dataset. The highest scores are highlighted in red.}
    \label{tab:rdst-hrl-detail}
    \centering
    \begin{tabular}{l|llll}
\hline
\textbf{OASIS}                       & \textbf{Dice-T$\uparrow$}                      & \textbf{Dice-G$\uparrow$}                     & \textbf{Dice-W$\uparrow$}                     & \textbf{Dice-CSF$\uparrow$}                   \\
\textbf{SwinIR}                      & 0.8888(0.0097)                                 & 0.8506(0.054)                                 & 0.8688(0.020)                                 & 0.8880(0.015)                                 \\
\textbf{RDST-E($\mathcal{L}_{HRL}$)} & 0.8899(0.0082)                                 & 0.8532(0.050)                                 & {\color[HTML]{E74025} \textbf{0.8693(0.021)}} & {\color[HTML]{E74025} \textbf{0.8890(0.013)}} \\
\textbf{RDST($\mathcal{L}_{HRL}$)}   & {\color[HTML]{E74025} \textbf{0.8906(0.0081)}} & {\color[HTML]{E74025} \textbf{0.8567(0.049)}} & {\color[HTML]{E74025} \textbf{0.8693(0.021)}} & 0.8885(0.013)                                 \\ \hline
\textbf{BraTS}                       & \textbf{Dice-WT$\uparrow$}                     & \textbf{Dice-ET$\uparrow$}                    & \textbf{Dice-TC$\uparrow$}                    & \textbf{}                                     \\
\textbf{SwinIR}                      & 0.7836(0.12)                                  & 0.6816(0.12)                                  & 0.6710(0.16)                                  &                                               \\
\textbf{RDST-E($\mathcal{L}_{HRL}$)} & 0.7825(0.13)                                   & {\color[HTML]{E74025} \textbf{0.7039(0.12)}} & {\color[HTML]{E74025} \textbf{0.6922(0.17)}} &                                               \\
\textbf{RDST($\mathcal{L}_{HRL}$)}   & {\color[HTML]{E74025} \textbf{0.7842(0.13)}}  & 0.6970(0.12)                                  & 0.6869(0.17)                                  &                                               \\ \hline
\textbf{ACDC}                        & \textbf{Dice-T$\uparrow$}                      & \textbf{Dice-LV$\uparrow$}                    & \textbf{Dice-RV$\uparrow$}                    & \textbf{Dice-MC$\uparrow$}                    \\
\textbf{SwinIR}                      & 0.8705(0.026)                                  & 0.9001(0.043)                                 & 0.6964(0.12)                                  & 0.8456(0.039)                                 \\
\textbf{RDST-E($\mathcal{L}_{HRL}$)} & {\color[HTML]{E74025} \textbf{0.8745(0.024)}}  & {\color[HTML]{E74025} \textbf{0.9005(0.044)}} & 0.7068(0.12)                                  & {\color[HTML]{E74025} \textbf{0.8501(0.032)}} \\
\textbf{RDST($\mathcal{L}_{HRL}$)}   & 0.8732(0.024)                                  & 0.8996(0.036)                                 & {\color[HTML]{E74025} \textbf{0.7197(0.11)}}  & 0.8476(0.036)                                 \\ \hline
\textbf{COVID-CT}                    & \textbf{Dice-T$\uparrow$}                      & \textbf{Dice-LL$\uparrow$}                    & \textbf{Dice-RL$\uparrow$}                    & \textbf{Dice-Lesion$\uparrow$}                \\
\textbf{RCAN}                        & 0.8365(0.17)                                   & 0.8175(0.21)                                  & {\color[HTML]{E74025} \textbf{0.8684(0.13)}}  & {\color[HTML]{E74025} \textbf{0.6207(0.11)}}  \\
\textbf{RDST-E($\mathcal{L}_{HRL}$)} & 0.8347(0.18)                                   & 0.8237(0.21)                                  & 0.8632(0.14)                                  & 0.5974(0.097)                                 \\
\textbf{RDST($\mathcal{L}_{HRL}$)}   & {\color[HTML]{E74025} \textbf{0.8411(0.16)}}   & {\color[HTML]{E74025} \textbf{0.8265(0.20)}}  & 0.8585(0.15)                                  & 0.6173(0.089)                                 \\ \hline
\end{tabular}
    
\end{table}

\noindent \textbf{SR for segmentation} we fine-tune RDST and RDST-E with $\mathcal{L}_{HRL}$ and both models achieve superior segmentation performance than SOTA methods on all datasets (Table. \ref{tab:rdst-hrl}). Compared with $\mathcal{L}_{E(1)}$ fine-tuned RDST variants, the RDST-E (with $\mathcal{L}_{HRL}$) improves the segmentation performance of the whole regions in the experiments with the OASIS, the ACDC and the COVID-CT datasets and increases the dice coefficient scores by 0.0040 on average of all four datasets, while the RDST (with $\mathcal{L}_{HRL}$) improves the segmentation performance on the OASIS, the BraTS and the ACDC datasets and increases the dice coefficient scores by 0.0008 on average of all four datasets. In addition, we choose one SOTA method with the best segmentation performance for each dataset and compare it with the $\mathcal{L}_{HRL}$ fine-tuned RDST variants in detail (Table. \ref{tab:rdst-hrl-detail}). Among the 15 sub-regions in total, both RDST and RDST-E (with $\mathcal{L}_{HRL}$) achieve the best scores on 7 sub-regions (with one overlap) and RCAN achieves the best segmentation performance of the right lung and lesion of the COVID-CT dataset.
\\

\begin{figure*}[ht]
    \centering
    \includegraphics[width=0.9\textwidth]{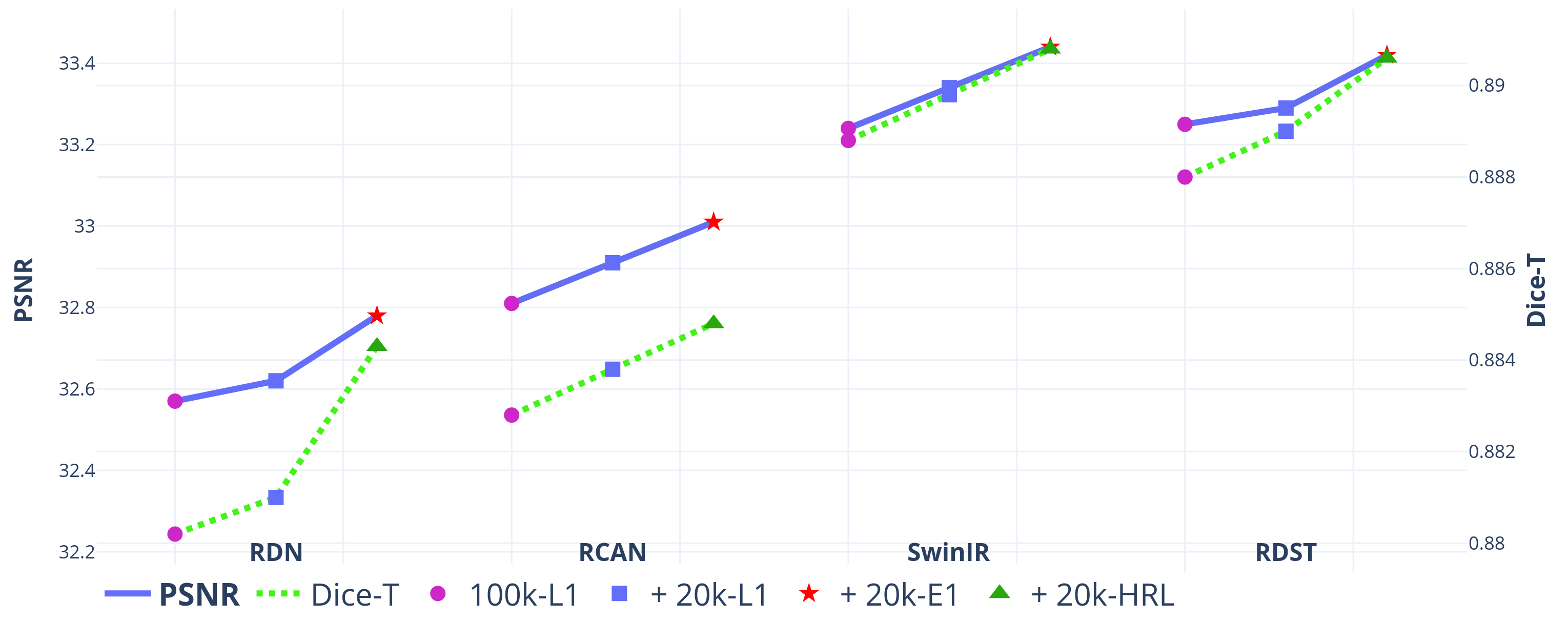}
    \caption{Extend the segmentation based perceptual loss variations $\mathcal{L}_{E(1)}$ and $\mathcal{L}_{HRL}$ to three popular SOTA methods: RDN \cite{RDN}, RCAN \cite{RCAN} and SwinIR \cite{swinir}. For each method, three fine-tuned variations (with $\mathcal{L}_{1}$, $\mathcal{L}_{E(1)}$ and $\mathcal{L}_{HRL}$, respectively) are compared with the baseline (trained for 100k steps with $\mathcal{L}_{1}$ only).} 
    \label{fig:rdst_sota_seg}
\end{figure*}

\noindent \textbf{Extending to SOTA methods} The proposed segmentation-based perceptual loss variations can successfully extend to other SOTA methods (Tabel. \ref{tab:rdst-seg-sota} and Fig. \ref{fig:rdst_sota_seg}). To verify the universality and usability, $\mathcal{L}_{E{1}}$ and $\mathcal{L}_{HRL}$ are used to fine-tune three popular SOTA SISR methods, including CNNs and vision transformers: RDN [\cite{RDN}], RCAN [\cite{RCAN}] and SwinIR [\cite{swinir}]. In contrast to the baselines (trained with $\mathcal{L}_{1}$ for 100k steps), extra training steps with only $\mathcal{L}_{1}$ bring limited improvements of PSNR (+0.04 dB) and segmentation performance (+0.0005 Dice-T/WT) on average. Conversely, the proposed segmentation-based perceptual loss variations significantly boost both image quality (+0.16 dB PSNR on average with $\mathcal{L}_{E(1)}$) and segmentation accuracy (+0.0028 Dice-T/WT on average with $\mathcal{L}_{HRL}$). Similar to the above conclusion in this section, models fined-tuned with $\mathcal{L}_{HRL}$ achieve the best dice scores in the downstream segmentation tasks in all cases, as expected. On the other hand, models fine-tuned with $\mathcal{L}_{E1}$ achieve the best PSNR for all cases and the second-best dice scores in most cases.
\\

\noindent \textbf{To datasets without segmentation labels} In the above experiments, the segmentation-based perceptual loss $\mathcal{L}_{E(1)}$ has demonstrated its robust applicability and effectiveness. In most cases, it significantly improves image fidelity quality (i.e. PSNR and SSIM) with comparable segmentation performance. Although we train a segmentation model for each dataset independently, the training is not a limitation. To use $\mathcal{L}_{E(1)}$, a U-Net can be trained on one dataset with segmentation labels and straightly extended to a new dataset without segmentation labels. In the simulation experiment, we use the pre-trained U-Net models with the OASIS, the ACDC and the COVID-CT datasets to fine-tune the RDST-E model of every dataset and achieve equal performance (Table. \ref{tab:rdst_seg_loss_tl}). Compared with the baselines (120k steps training with $\mathcal{L}_{1}$), they result in obvious improvement of PSNR and dice coefficient scores in almost all cases. The prior knowledge of the pre-trained segmentation model with one medical image dataset can be effectively transferred to new datasets. Proposed segmentation perceptual loss $\mathcal{L}_{E(1)}$ can become regular in a wider range of medical image low-level tasks.
\\

\subsection{Limitations and future works}\label{sec:limitations}
\noindent Although the above experiments illustrate the superior performance and robustness of the proposed method, three limitations are worth to be noticed. First of all, all results and comparisons are achieved in simulation SR tasks. The degradation and noise formulation we used for HR-LR image pair generation in Section \ref{sec:experiments_data} may not represent the actual condition of various medical modalities in the clinic. Thus, it is worth exploring the capacity of the proposed method in enhancement tasks with clinical medical images in the future. Second, it is challenging to avoid over-fitting in medical image SR tasks. In Section \ref{sec:result_sota}, the proposed method RDST achieves the best performance (i.e. PSNR) for three datasets but achieves worse performance than the lite version RDST-E with the small dataset ACDC. we deduce that the decline is caused by over-fitting because fewer training steps of regular-size models (e.g. RDST and SwinIR) result in higher PSNR scores in the experiments of ACDC. Thus, developing data-driven early-stopping methods [\cite{ying2019overfitting}] for each case of medical image dataset is necessary and significant. Third, in the ablation study of model efficiency (Section \ref{sec:result_efficiency}), vision transformers are much slower in inference than CNN models with similar sizes. Although RDST variants have achieved the smallest model size and fewest parameters, non-attention CNN methods (i.e. EDSR and RDN) are still more than twice faster as our proposed method. Exploring more efficient vision transformers [\cite{rao2021efficient_ViT, tang2022efficient_ViT}] with the remaining SR performance will be very interesting. 
\\
\\
\noindent Additional feature works can also be arranged in the following two directions. On the one hand, the proposed method can be a potential backbone for broader low-level medical image analysis tasks. For example, the residual dense vision transformer can be extended to MR and CT synthesis [\cite{wolterink2017CT_MR_synthesis}] tasks with shallow feature extraction and up-sampler layers modifications. Meanwhile, the proposed perceptual losses can be alternatives in MR imaging reconstruction and image registration [\cite{xue2022MR_reconstruction, han2022deformable_MR_registration}]. On the other hand, super-resolution tasks can be integrated into more downstream medical image analysis tasks than segmentation. Thus, the proposed method can be introduced to more medical modalities in addition to radiology scans. For example, novel perceptual losses may be designed with pre-trained models of retinal image classification [\cite{playout2022Retinal_image_classification}] and improve the performance of retinal image synthesis [\cite{zhao2018Retinal_image_synthesis}].

\begin{table*}[t]
\caption{Extending the proposed segmentation-based perceptual losses to SOTA methods. For each method, the baseline model is trained with $\mathcal{L}_{1}$ only for 100k steps. Three fine-tuned variations are additionally trained for 20k steps with: (1) $\mathcal{L}_{1}$ only; (2) $\mathcal{L}_{1}$ with $\mathcal{L}_{E(1)}$; and (3) $\mathcal{L}_{1}$ with $\mathcal{L}_{HRL}$. In each group of the same method, the best and the second-best scores are in highlighted in red and blue, respectively.}
\label{tab:rdst-seg-sota}
\centering
\begin{tabular}{ll|llllll}
\hline
\multicolumn{2}{c|}{\textbf{Model Training}}                   & \textbf{PSNR$\uparrow$}                    & \textbf{SSIM$\uparrow$}                       & \textbf{Dice-T$\uparrow$}                      & \textbf{Dice-G$\uparrow$}                     & \textbf{Dice-W$\uparrow$}                     & \textbf{Dice-CSF$\uparrow$}                   \\ \hline
                                  & 100k-$\mathcal{L}_{1}$     & 32.57(3.2)                                 & {\color[HTML]{E74025} \textbf{0.9241(0.036)}} & 0.8802(0.010)                                  & 0.8316(0.059)                                 & 0.8620(0.021)                                 & {\color[HTML]{326FBA} \textbf{0.8845(0.015)}} \\
                                  & +20k-$\mathcal{L}_{1}$     & {\color[HTML]{326FBA} \textbf{32.62(3.3)}} & 0.9217(0.037)                                 & 0.8810(0.010)                                  & 0.8343(0.058)                                 & 0.8621(0.021)                                 & 0.8840(0.015)                                 \\
                                  & +20k-$\mathcal{L}_{E(1)}$ & {\color[HTML]{E74025} \textbf{32.78(3.4)}} & {\color[HTML]{326FBA} \textbf{0.9227(0.037)}} & {\color[HTML]{326FBA} \textbf{0.8815(0.0099)}} & {\color[HTML]{326FBA} \textbf{0.8361(0.058)}} & {\color[HTML]{326FBA} \textbf{0.8622(0.021)}} & 0.8830(0.014)                                 \\
\multirow{-4}{*}{\textbf{RDN}}    & +20k-$\mathcal{L}_{HRL}$   & 32.30(3.1)                                 & 0.9153(0.038)                                 & {\color[HTML]{E74025} \textbf{0.8843(0.0088)}} & {\color[HTML]{E74025} \textbf{0.8440(0.052)}} & {\color[HTML]{E74025} \textbf{0.8640(0.021)}} & {\color[HTML]{E74025} \textbf{0.8859(0.014)}} \\ \hline
                                  & 100k-$\mathcal{L}_{1}$     & {\color[HTML]{326FBA} \textbf{32.81(3.6)}} & {\color[HTML]{326FBA} \textbf{0.9224(0.038)}} & 0.8828(0.0094)                                 & 0.8424(0.054)                                 & 0.8619(0.021)                                 & 0.8831(0.013)                                 \\
                                  & +20k-$\mathcal{L}_{1}$     & {\color[HTML]{326FBA} \textbf{32.81(3.6)}} & 0.9221(0.038)                                 & 0.8827(0.0094)                                 & 0.8419(0.055)                                 & 0.8619(0.021)                                 & 0.8828(0.013)                                 \\
                                  & +20k-$\mathcal{L}_{E(1)}$ & {\color[HTML]{E74025} \textbf{32.94(3.7)}} & {\color[HTML]{E74025} \textbf{0.9231(0.038)}} & {\color[HTML]{326FBA} \textbf{0.8833(0.0093)}} & {\color[HTML]{326FBA} \textbf{0.8432(0.054)}} & {\color[HTML]{326FBA} \textbf{0.8623(0.021)}} & {\color[HTML]{326FBA} \textbf{0.8836(0.014)}} \\
\multirow{-4}{*}{\textbf{RCAN}}   & +20k-$\mathcal{L}_{HRL}$   & 32.64(3.5)                                 & 0.9187(0.038)                                 & {\color[HTML]{E74025} \textbf{0.8851(0.0082)}} & {\color[HTML]{E74025} \textbf{0.8477(0.050)}} & {\color[HTML]{E74025} \textbf{0.8637(0.021)}} & {\color[HTML]{E74025} \textbf{0.8841(0.013)}} \\ \hline
                                  & 100k-$\mathcal{L}_{1}$     & 33.24(3.7)                                 & 0.9287(0.036)                                 & 0.8888(0.0097)                                 & 0.8506(0.054)                                 & 0.8688(0.020)                                 & {\color[HTML]{326FBA} \textbf{0.8880(0.015)}} \\
                                  & +20k-$\mathcal{L}_{1}$     & {\color[HTML]{326FBA} \textbf{33.33(3.7)}} & {\color[HTML]{326FBA} \textbf{0.9295(0.036)}} & 0.8891(0.010)                                  & {\color[HTML]{326FBA} \textbf{0.8523(0.054)}} & 0.8688(0.021)                                 & 0.8872(0.015)                                 \\
                                  & +20k-$\mathcal{L}_{E(1)}$ & {\color[HTML]{E74025} \textbf{33.46(3.7)}} & {\color[HTML]{E74025} \textbf{0.9303(0.036)}} & {\color[HTML]{326FBA} \textbf{0.8893(0.011)}}  & 0.8521(0.054)                                 & {\color[HTML]{326FBA} \textbf{0.8692(0.021)}} & 0.8874(0.015)                                 \\
\multirow{-4}{*}{\textbf{SwinIR}} & +20k-$\mathcal{L}_{HRL}$   & 33.08(3.6)                                 & 0.9248(0.037)                                 & {\color[HTML]{E74025} \textbf{0.8908(0.0086)}} & {\color[HTML]{E74025} \textbf{0.8573(0.048)}} & {\color[HTML]{E74025} \textbf{0.8697(0.021)}} & {\color[HTML]{E74025} \textbf{0.8899(0.014)}} \\ \hline
                                  & 100k-$\mathcal{L}_{1}$     & 33.25(3.5)                                 & 0.9286(0.035)                                 & 0.8880(0.0097)                                 & 0.8489(0.055)                                 & 0.8679(0.021)                                 & {\color[HTML]{2972F4} \textbf{0.8881(0.015)}} \\
                                  & +20k-$\mathcal{L}_{1}$     & {\color[HTML]{2972F4} \textbf{33.29(3.6)}} & {\color[HTML]{2972F4} \textbf{0.9293(0.035)}} & {\color[HTML]{2972F4} \textbf{0.8890(0.0094)}} & 0.8512(0.054)                                 & {\color[HTML]{2972F4} \textbf{0.8690(0.020)}} & 0.8877(0.015)                                 \\
                                  & +20k-$\mathcal{L}_{E(1)}$ & {\color[HTML]{E74025} \textbf{33.42(3.7)}} & {\color[HTML]{E74025} \textbf{0.9299(0.035)}} & 0.8889(0.0097)                                 & {\color[HTML]{2972F4} \textbf{0.8514(0.054)}} & 0.8688(0.021)                                 & 0.8874(0.015)                                 \\
\multirow{-4}{*}{\textbf{RDST}}   & +20k-$\mathcal{L}_{HRL}$   & 33.01(3.5)                                 & 0.9243(0.037)                                 & {\color[HTML]{E74025} \textbf{0.8906(0.0081)}} & {\color[HTML]{E74025} \textbf{0.8567(0.049)}} & {\color[HTML]{E74025} \textbf{0.8693(0.021)}} & {\color[HTML]{E74025} \textbf{0.8885(0.013)}} \\ \hline
\end{tabular}
\end{table*}

\begin{table*}[h]
\centering
\caption{A transfer learning study on the segmentation-based perceptual loss in the fine-tuning stage. For each testing dataset of OASIS, ACDC and COVID, the RDST is trained for 120k steps with only $\mathcal{L}_{1}$ as the baselines to avoid the impacts of extra training steps. In the fine-tuning stage, pre-trained U-Net models with OASIS, ACDC and COVID datasets are used respectively to calculate $\mathcal{L}_{E(1)}$. Scores in red indicate better performance than baselines, and scores in green indicate worse performance.}
\label{tab:rdst_seg_loss_tl}
\begin{tabular}{l|l|lll}
\hline
\textbf{Testing}     & \textbf{Baseline}      & \multicolumn{3}{c}{\textbf{Which U-Net for $\mathcal{L}_{E(1)}$}}                                                                             \\ \hline
\textbf{PSNR$\uparrow$} & \textbf{120k-$\mathcal{L}_{1}$}  & \textbf{OASIS}                        & \textbf{ACDC}                        & \textbf{COVID}                        \\
OASIS                       & 33.12(3.4)     & {\color[HTML]{E74025} 33.26(3.4)}     & {\color[HTML]{E74025} 33.27(3.4)}    & {\color[HTML]{E74025} 33.26(3.4)}     \\
ACDC                        & 27.23(3.0)     & {\color[HTML]{E74025} 27.24(3.0)}     & {\color[HTML]{E74025} 27.24(3.0)}    & {\color[HTML]{E74025} 27.24(3.0)}     \\
COVID              & 34.58(4.4)     & {\color[HTML]{E74025} 34.60(4.4)}     & {\color[HTML]{E74025} 34.62(4.4)}    & {\color[HTML]{E74025} 34.62(4.5)}     \\ \hline
\textbf{Dice-T$\uparrow$}   & \textbf{120k-$\mathcal{L}_{1}$}  & \textbf{OASIS}                        & \textbf{ACDC}                        & \textbf{COVID}                        \\
OASIS                       & 0.8874(0.0094) & {\color[HTML]{53AD5B} 0.8871(0.0096)} & 0.8874(0.0094)                       & {\color[HTML]{E74025} 0.8875(0.0094)} \\
ACDC                        & 0.8681(0.026)  & {\color[HTML]{E74025} 0.8684(0.025)}  & {\color[HTML]{E74025} 0.8691(0.025)} & {\color[HTML]{E74025} 0.8683(0.025)}  \\
COVID                       & 0.8241(0.19)   & {\color[HTML]{E74025} 0.8284(0.18)}   & {\color[HTML]{E74025} 0.8301(0.18)}  & {\color[HTML]{E74025} 0.8264(0.19)}   \\ \hline
\end{tabular}
\end{table*}

\section{Conclusion}\label{sec:conclusion}
\noindent In this work, we aim to improve the single-image super-resolution performance and efficiency of supervised vision transformers on medical images by introducing popular mechanisms in previous CNNs to transformers and transferring prior knowledge of segmentation tasks to SR tasks. We present an efficient and robust single-image super-resolution method for medical images by successfully introducing the residual dense connection and local feature fusion to vision transformers. This developed RDST and its efficient version RDST-E have achieved superior or equal performance to the SOTA SISR methods of both SR image fidelity quality and downstream auto segmentation tasks. In the simulation experiments of four public medical image datasets, including MR and CT scans, the proposed approaches have resulted in averaged improvements of +0.09 dB and +0.06 dB PSNR receptively with only 38\% and 20\% parameters of SwinIR. Meanwhile, we implement a perceptual loss for SR tasks based on the prior knowledge of pre-trained segmentation models and successfully extend its variants to SOTA methods, including CNNs and ViTs. The perceptual loss variant for reconstruction fidelity has led to an improvement of +0.14 dB PSNR on average, and the variant for machine perception (i.e. downstream segmentation tasks) has led to an improvement of 0.0023 dice coefficient on average. In summary, this work has introduced a framework with novel and practical designs on model architecture, loss function, training tricks and evaluation metrics. It has achieved SOTA performance in super-resolution tasks with various medical image modalities. It is also a potential backbone for more medical image low-level tasks such as reconstruction and synthesis.

\section*{Acknowledgments}
\noindent Jin Zhu was supported by China Scholarship Council (201708060173). Guang Yang was supported in part by the BHF (TG/18/5/34111, PG/16/78/32402), the ERC IMI (101005122), the H2020 (952172), the MRC (MC/PC/21013), the Royal Society (IEC/NSFC/211235), and the UKRI Future Leaders Fellowship (MR/V023799/1).


\bibliographystyle{model2-names.bst}\biboptions{authoryear}
\bibliography{refs}



\end{document}